\documentclass[runningheads]{llncs}
\newcommand*{\ARXIV}{}%

\usepackage[hidelinks]{hyperref} 
\usepackage{graphicx} 
\usepackage{amsfonts} 
\usepackage[bottom]{footmisc} 
\usepackage{subfig}
\usepackage{verbatim} 
\usepackage{enumitem} 
\usepackage{times}
\usepackage{algorithm}
\usepackage[noend]{algpseudocode}

\makeatletter
\def\BState{\State\hskip-\ALG@thistlm}
\makeatother

\begin{document}
\pagestyle{headings}

\title{Algorithms for Stable Matching and\\ Clustering in a Grid}
\titlerunning{Stable Matching and Clustering in a Grid}
\author{David Eppstein \and Michael T. Goodrich \and
Nil Mamano}
\authorrunning{D. Eppstein \and M.T. Goodrich \and N. Mamano}
\institute{Department of Computer Science, University of California, Irvine, USA\\
\email{eppstein@uci.edu}, \texttt{goodrich@acm.org}, \texttt{nmamano@uci.edu}}


\maketitle             

\begin{abstract}
We study a discrete version of a geometric stable marriage problem
originally proposed in a continuous setting by Hoffman, Holroyd,
and Peres, in which points in the plane are stably matched to cluster
centers, as prioritized by their distances, so that each
cluster center is apportioned a set of points of equal area. We show
that, for a discretization of the problem to an $n\times n$ grid
of pixels with $k$ centers, the problem can be solved in time
$O(n^2 \log^5 n)$, and we experiment with two slower but more
practical algorithms and a hybrid method that switches
from one of these algorithms to the other to gain greater efficiency
than either algorithm alone. We also show how to combine geometric stable
matchings with a $k$-means clustering algorithm, so as to 
provide a geometric political-districting algorithm that 
views distance in economic terms, and we experiment
with weighted versions of stable $k$-means in order to improve the
connectivity of the resulting clusters.
\end{abstract}

\section{Introduction}
\begin{figure}[hbt]
\centering
\ifdefined\ARXIV
\includegraphics[width=0.7\linewidth]{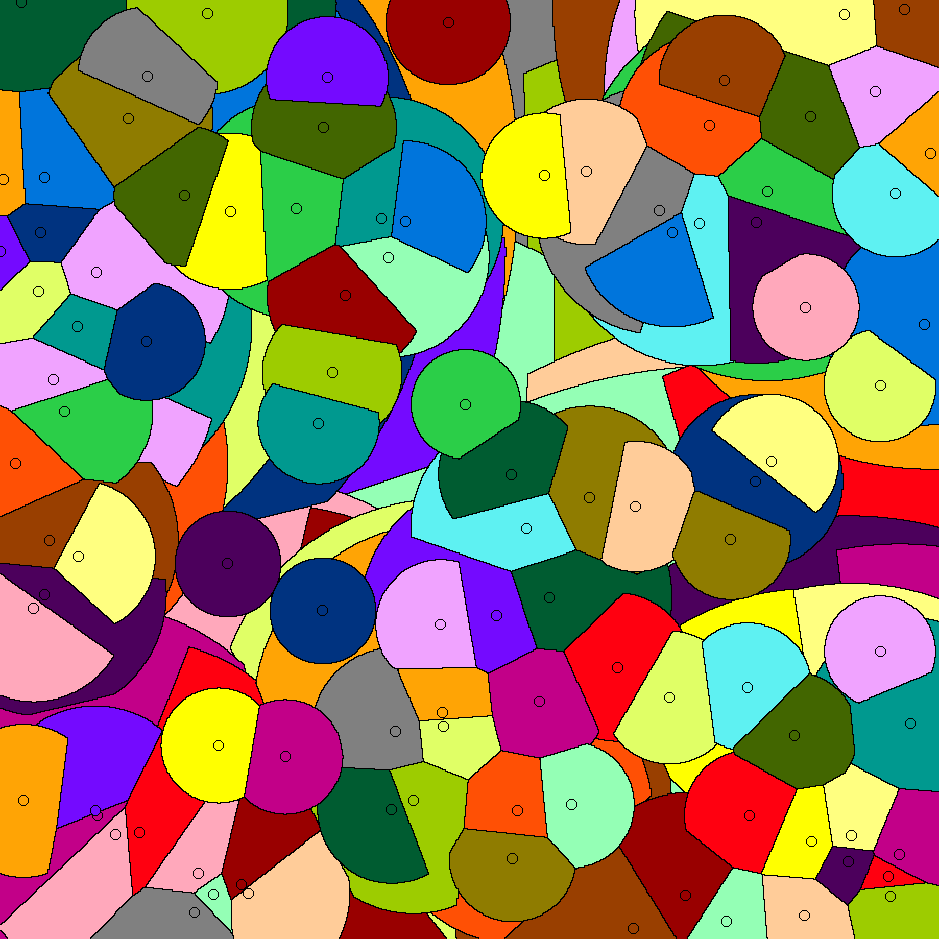}
\else
\includegraphics[width=0.5\linewidth]{figure1}
\fi
\caption{An example solution to 
the \emph{stable grid matching} problem for a $900\times900$ grid and
$100$ centers distributed randomly. Pixels of the same color are assigned to the same center.}
\label{fig:example}
\end{figure}

A long line of research considers algorithms
on objects embedded in $n\times n$ grids,
including problems in 
computational geometry
(e.g., see~\cite{AKMAN1989410,ARKIN200025,Chan09,fang1993delaunay,Greene86,keil1980computational,Overmars1988,OVERMARS1988254}),
graph drawing
(e.g., see~\cite{Biedl2013,CHROBAK199829,DeFraysseix1990,RAHMAN1998203}),
geographic information systems
(e.g., see~\cite{de1999applications}),
and
geometric image processing
(e.g., see~\cite{Chandran1992,Chun2009,Dehne1990,hartley2003multiple}).
Continuing this line, 
we consider in this paper the problem of matching grid points 
(which we view as \emph{pixels})
to $k$ \emph{center} points in the grid.
Pixels have a preference for centers closer to them, and centers prefer
closer pixels as well. The goal is to match
every center to an equal number of pixels and for the matching
to be \textit{stable}, meaning that no two elements prefer each
other to their specified matches.
For example, the centers could be facilities, such as polling places,
fire stations, or post offices, that have assigned jurisdictions and
equal operational capacities (in terms of how many pixels they can
serve). Rather than optimizing some
computationally challenging global quality criterion based on distance or area, 
we seek an assignment of pixels to centers that is 
locally stable.
Figure~\ref{fig:example} illustrates a solution to this 
\emph{stable grid matching} problem for a $900\times 900$ grid 
and 100 random centers.
Note that some centers are matched to disconnected regions.  

Stable grid matching is a special case of the 
classic \textit{stable matching problem}~\cite{gale62}, which was originally
described in terms of arranging marriages between $N$ heterosexual
men and women in a closed community. 
In this case, stability means that no man-woman pair prefers each
other to their assigned mates, which is necessary
(and more important than, e.g., total utility) to prevent
extramarital affairs. The Gale-Shapley algorithm~\cite{gale62} finds a
stable matching for arbitrary preferences in $O(N^2)$ time.
For stable grid matching in an $n\times n$ grid
this would give a running time of $O(n^4)$, since each ``man'' would correspond
to a pixel and each ``woman'' would correspond to
one of $\lceil n^2/k\rceil$ copies of a center. As we show, 
the geometric structure of the stable grid matching
problem allows for significantly more efficient solutions.

We also study the effect of integrating a stable matching with a
\textit{k-means} clustering method, which alternates between assigning
points to cluster centers and moving cluster centers to better
represent their assigned points. Using stable matching for the
assignment stage of this method allows us to fix the size of the
clusters (for instance, to be all equally sized), which might be
advantageous in some applications.

\paragraph{Prior Related Work.}
As mentioned above, there is considerable prior research on algorithms
involving objects embedded in an $n\times n$ grid.
The stable grid matching problem that we study can be viewed as 
a grid-restricted version of the classic ``post office'' problem
of Knuth~\cite{knuth1998art}, where one wishes to identify each point in
the plane with its closest of $k$ post offices, with the added restriction that 
the region assigned to each post office must have the same area.
The continuous version of the stable grid matching problem, which deals with
points in $\mathbb{R}^2$ instead of discrete pixels, was studied
by Hoffman et~al.~\cite{hoffman2006}. 
They showed that there is a unique solution,
and there is a simple numerical method to find it: Start growing a circle
from each center at the same time, all growing at the same speed.
When a yet-unmatched point is reached by a circle, it is assigned
to the corresponding center. When a center reaches its quota (its
region covers $1/k$ of the area of the square), its circle halts.
(Note that if the halting condition is removed, we obtain the Voronoi
diagram of the centers instead, as in the well-known solution
to Knuth's post office problem, e.g., see~\cite{Aurenhammer:1991}.)
Due to its continuous, numerical nature, Hoffman et al.~did not analyze
the running time of their method; hence, there is motivation
to study the grid-based version of this problem.

With respect to the related problem of $k$-means clustering, we
are interested in a grid-based version of this problem as well, which
has been studied extensively in non-grid discrete contexts
(e.g., see~\cite{Kanungo02,Jain:1999}).
In the continuous version of this problem, one is interested in partitioning
a geometric region into subregions that all have the same area
(e.g., see~\cite{Bohringer1999}).
One of the motivations for such partitions is in
\textit{political districting},
for which there is additional related prior work
(e.g., see~\cite{solbrig2013}).
The goal of political districting is to partition a territory into
regions (districts) which all have roughly the same population size
and are ``compact'', which informally means that their shape should
be connected and resemble a circle rather than an
octopus~\cite{solbrig2013}. Ricca et~al.~\cite{Ricca2008voronoi} adapted the
concept of Voronoi regions to the discrete setting in order to use
them for political districting. Voronoi regions ensured good
compactness but poor population balance, however. 
Thus, there is motivation
for a clustering algorithm based on the use of 
stable matchings, since such partitions enforce the property that
all regions have the same size (at the possible cost of connectivity).
Finding a scheme that guarantees both size equality and compactness
is an open problem of interest.

\begin{figure} \centering
\begin{minipage}[b]{0.42\linewidth}
\includegraphics[width=0.99\linewidth]{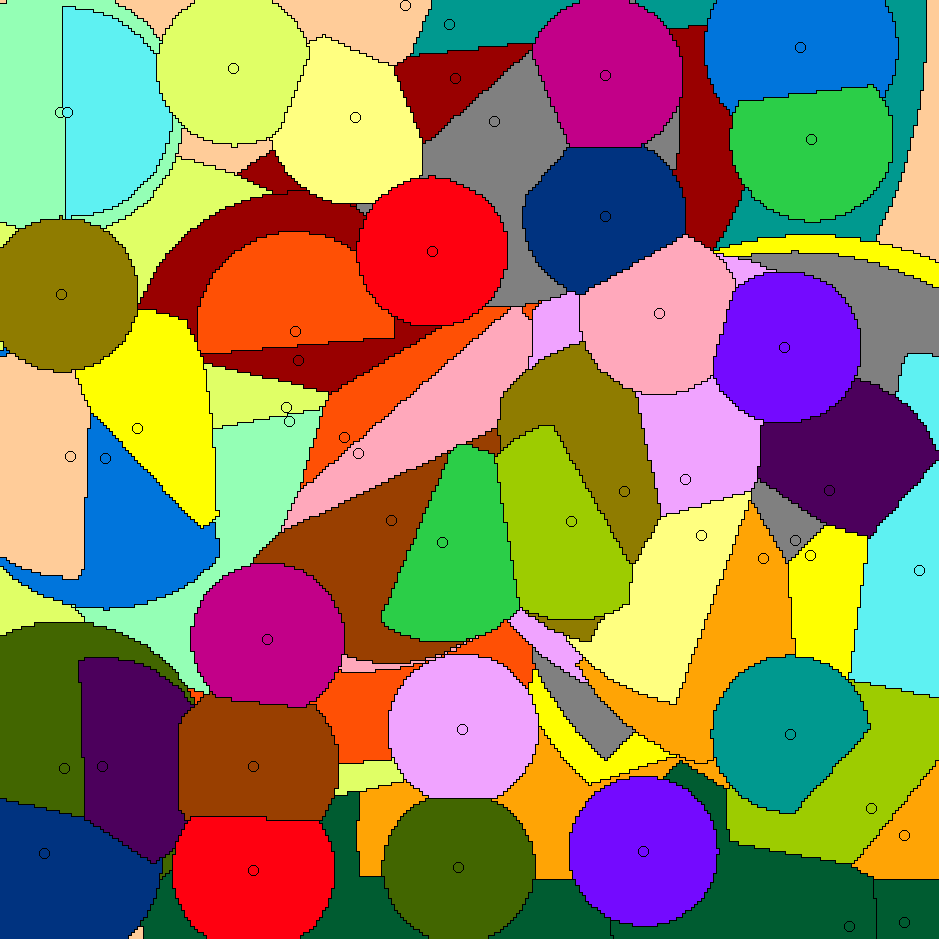}\\[4pt]
\includegraphics[width=0.99\linewidth]{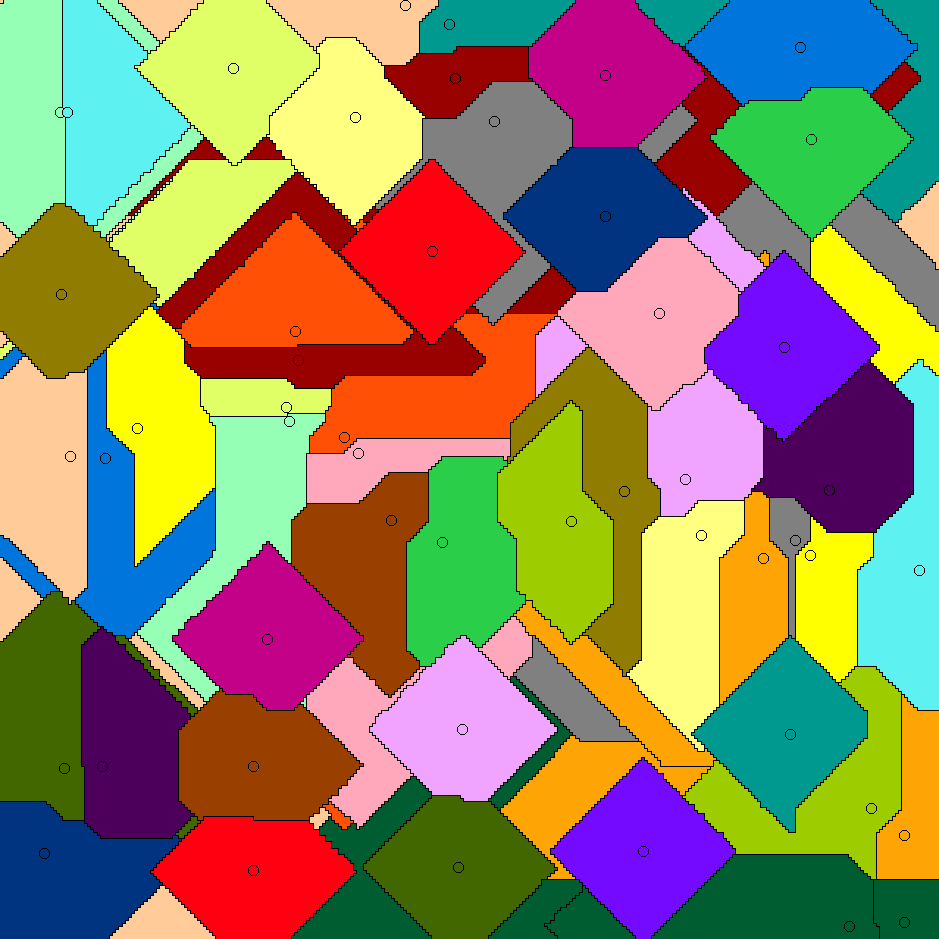}\\[4pt]
\includegraphics[width=0.99\linewidth]{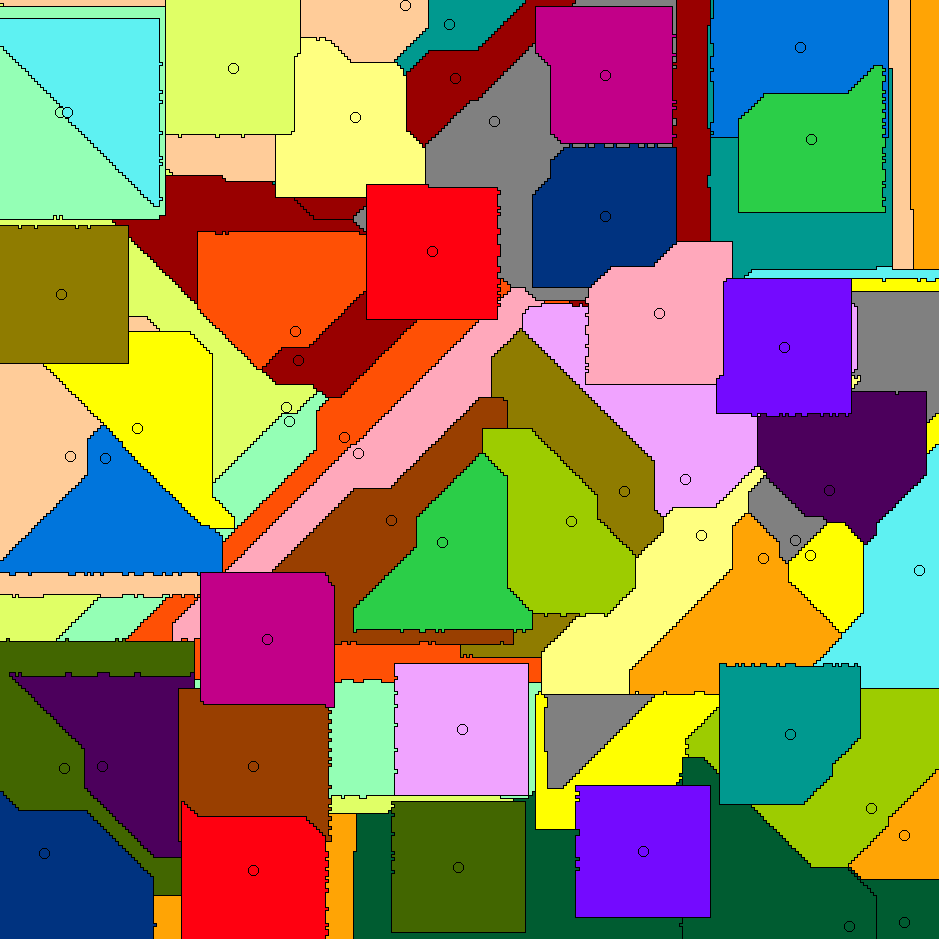}
\end{minipage} \begin{minipage}[b]{0.42\linewidth}
\includegraphics[width=0.99\linewidth]{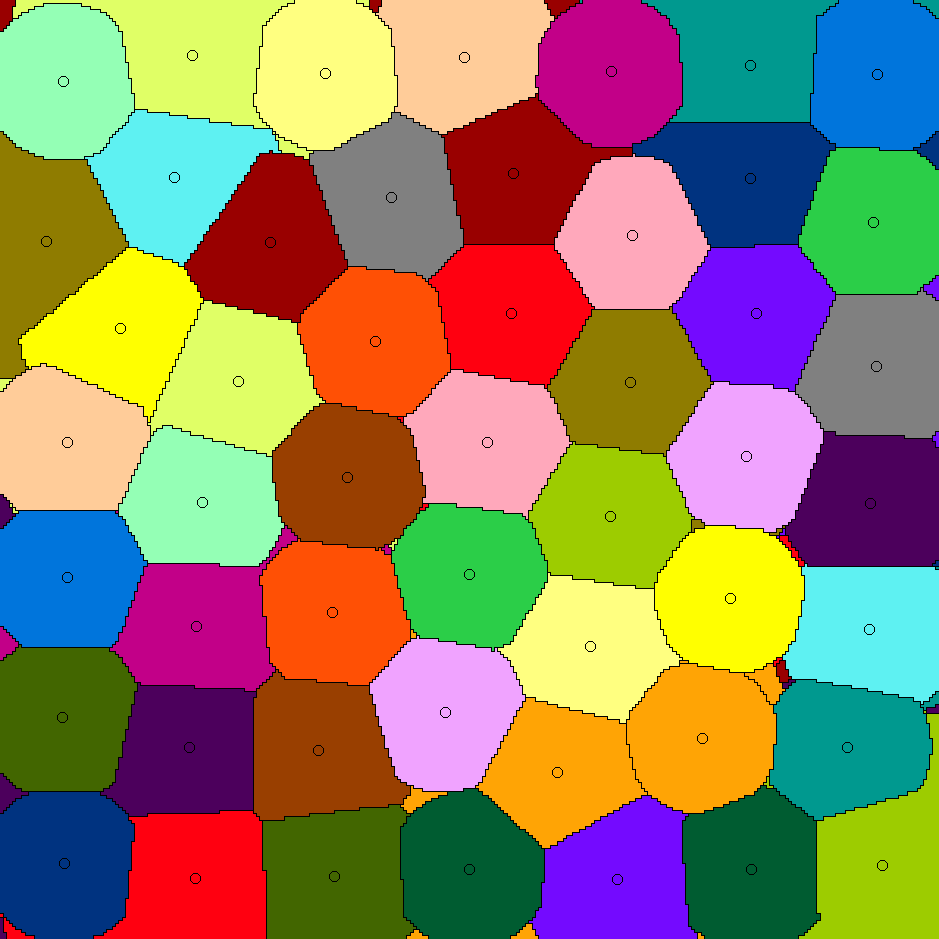}\\[4pt]
\includegraphics[width=0.99\linewidth]{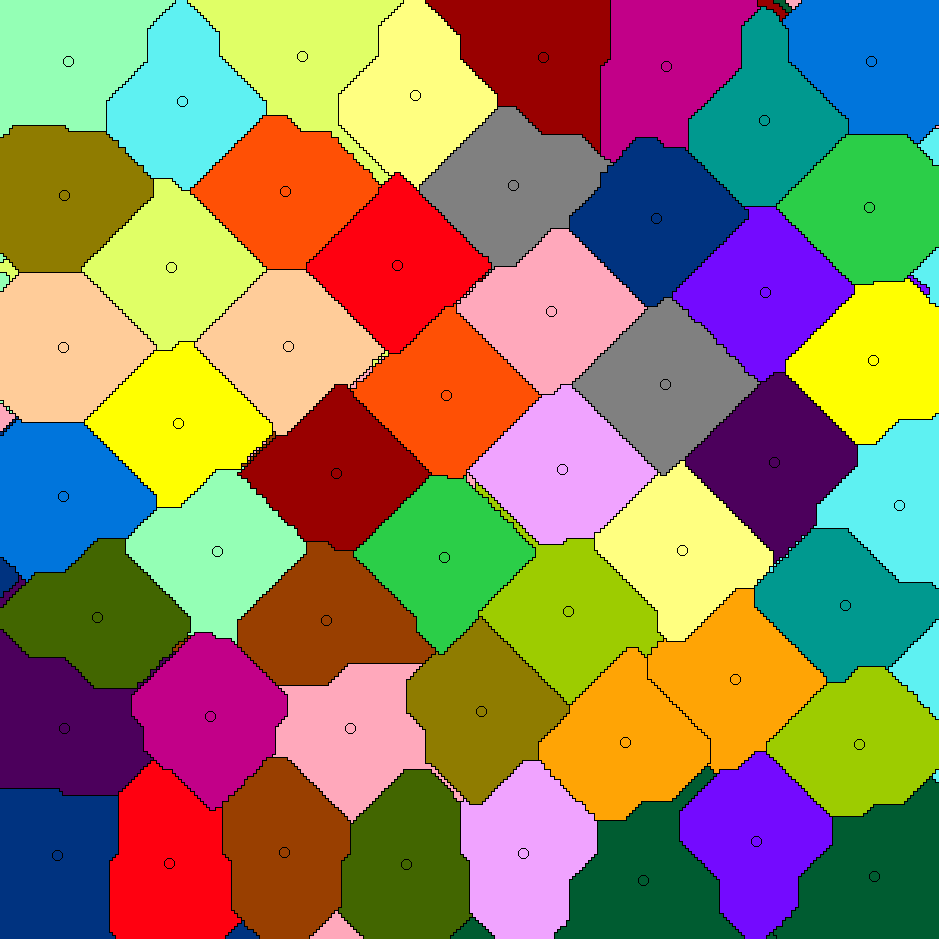}\\[4pt]
\includegraphics[width=0.99\linewidth]{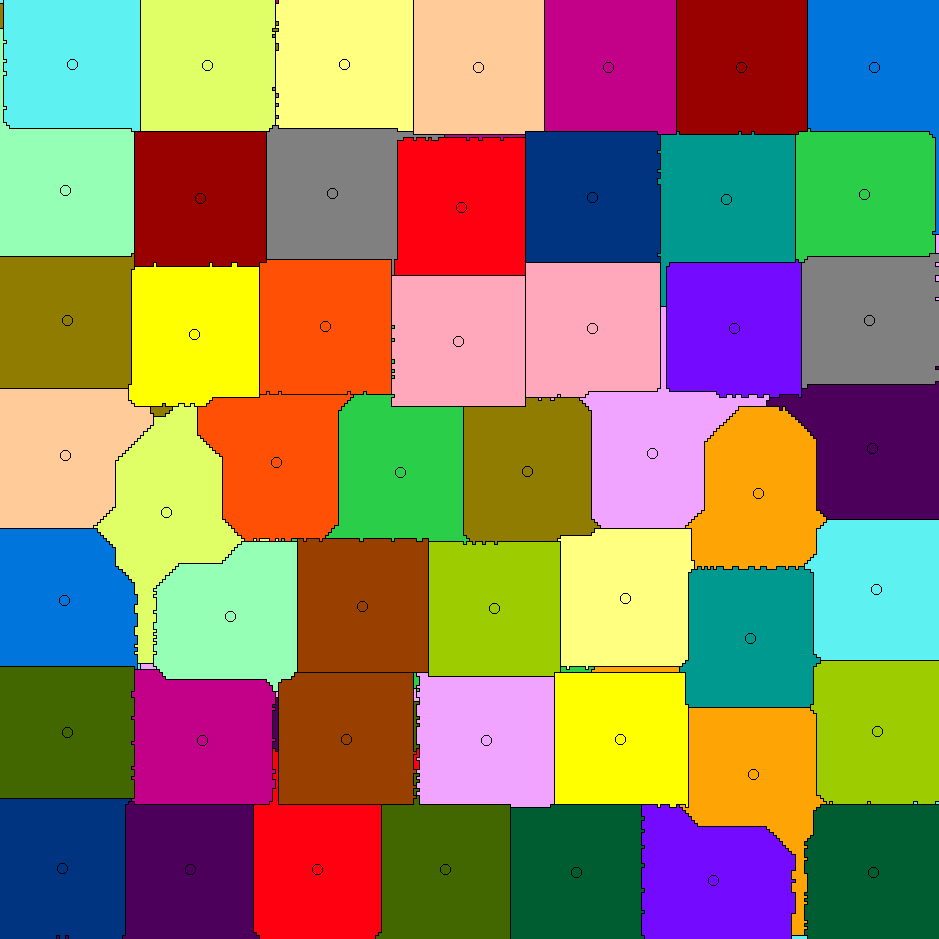}
\end{minipage} \caption{Left: stable matching in a $300\times300$
grid with the same $50$ random centers for the Euclidean (top),
Manhattan (center), and Chebyshev (bottom) metrics. Right: result
of the stable $k$-means algorithm with unweighted centroids for
each metric.} \label{fig:kmeans} 
\end{figure}

\paragraph{Problem Definition.}
In the \textit{stable grid matching problem}, we are given a square $n\times n$ grid and $k$ points called \textit{centers} within the grid. 
The lattice points are called \emph{pixels} or 
\emph{sites}. Sites implicitly rank the
centers in increasing order of distance, and centers similarly
implicitly rank pixels in increasing order by distance.
A \textit{matching} is a mapping from sites to
centers. The goal is to find a matching with the following two
properties (see Figure~\ref{fig:kmeans}, left column): 
\begin{enumerate} 
\item The \textit{region} of each
center (the set of sites assigned to it) must have the same size
up to roundoff errors. The \textit{quota} of a center is the number
of sites that must be in its region. If $n^2$ is a multiple of $k$,
then all the quotas are $n^2/k$. Otherwise, some centers are allowed
one extra site.  
\item The matching must be \textit{stable}. A
matching is not stable when a pair of sites $(p_1,p_2)$ is
assigned to centers $c_1$ and $c_2$ such that $p_1$ prefers (i.e., according
to some metric is closer to) 
$c_2$ over $c_1$ and $c_2$ prefers $p_1$ over $p_2$. This is unstable
because $p_1$ and $c_2$ prefer each other to their current matches.
\end{enumerate} 

\paragraph{Combining \textit{k}-means  with Stable Assignment.}
The \emph{$k$-means clustering} method is to 
partition a data set 
(which, in our case, is an $n\times n$ grid) into $k$ regions,
based on a simple iterative refinement algorithm (which is called
the \emph{$k$-means algorithm} or
\emph{Lloyd's algorithm}, e.g., see~\cite{Kanungo02}):
We begin by choosing
$k$ points, called cluster centers, randomly in the space. Then,
we iteratively repeat the following two phases: 1) \emph{assignment} step:
each object is assigned to its closest center, and 2) \emph{update} step:
each center is moved to the centroid of the objects assigned to it.

Lloyd's algorithm converges to a (locally optimal) partition that
minimizes the sum of the squared distances from each object to its
assigned center~\cite{Kanungo02}. 
In this paper, we propose a variation, which we 
call \textit{stable $k$-means}, 
where the assignment step is replaced by a stable matching
between objects and centers, so as to achieve the additional
property that the regions all have equal area (to within roundoff errors).  
Intuitively, the goal is to implement Lloyd's algorithm
with stable grid matching so as to improve the
compactness of the regions while preserving equal-sized clusters.

We have found through experimentation that,
although the stable $k$-means method succeeds in improving compactness, centers
can sometimes stop moving while we are executing Lloyd's algorithm
before their regions became completely connected
(e.g., see Figure~\ref{fig:kmeans}). 
Thus, we introduce in this paper an additional heuristic, where
we use weighted centroids, which are more sensitive
to the outlying parts of their region. The usual centroid of a set
of points $S$ is defined as $(\sum_{q\in S} q)/|S|$, where the
points are regarded as two-dimensional vectors so that the sum makes
sense. Instead, we can compute a weighted centroid as $(\sum_{q\in
S} w_q q)/(\sum_{q\in S} w_q)$. A natural choice to use for the
weight $w_q$ of a point $q$ assigned to the region of the center $c$ is
the distance from $q$ to $c$ raised to some exponent $p$ that we can choose, $d(q,c)^p$. The larger $p$
is, the more sensitive the weighted centroids are to outliers. When
$p=0$, we get the usual centroid. When $p\rightarrow+\infty$, we
get the circumcenter of the region, and when $p\rightarrow-\infty$
we get the current center.

\paragraph{Contributions.}
In this paper, we provide the following results:
\begin{itemize}
\item The stable  grid matching problem, for a grid of $n\times n$
pixels with $k$ centers, can be solved by a randomized algorithm
with expected running time $O(n^2\log^5 n)$.
Since an $n\times n$ grid has $\Theta(n^2)$ pixels, this quasilinear
bound improves the $O(n^4)$ time of the Gale-Shapley algorithm.
However, this algorithm uses intricate data structures
that make it challenging to implement in practice. 

\item 
Given the pragmatic challenges of the above-mentioned quasilinear-time 
algorithm, we provide two
alternative algorithms, a ``circle-growing algorithm''
and a ``distance-sorting'' method, both of which 
are simple to implement and have running times of $O(n^2 k)$.

\item 
We provide an experimental analysis of these two practical algorithms,
where
we observe that the circle-growing algorithm is more efficient at
finding low-distance matched pairs, while the distance-sorting based
method is more efficient when pairs are farther apart. Therefore,
we show that
it is advantageous to switch from one algorithm to the other partway
through the matching process, potentially achieving running times
with a sublinear dependence on $k$. We experiment with the optimal
cutoff for switching between these two algorithms.

\item We also provide the results of experiments to test the connectivity of the
clusters obtained by our stable $k$-means algorithm, 
with weighted variants for finding centroids.
Our experiments support the conclusion that
no choice of a weight exponent $p$
will always result in total connectivity. 
Nevertheless, our experiments provide evidence that
the best results come from the range
$-0.8\leq p\leq0.4$. 
Empirically, more highly negative values of $p$ tend to make the
algorithm converge slowly or fail to converge, while more highly
positive values of $p$ lead to oscillations in the center placement.
\ifdefined\ARXIV
See Appendix~\ref{app:centroids} for additional figures of these cases.
\else
See the full version of the paper for additional figures of these
cases.
\fi
\end{itemize}

\section{Algorithms}

Our stable grid matching algorithms start with an empty matching and add center--site pairs to it. Given a partial matching, we say a site is \textit{available} if it has not been matched yet, and a center is \textit{available} if the size of its region is smaller than its quota. A center--site pair is available if both the center and site are available, and it is a \textit{closest available pair} if it is available and the distance from the center to the site is minimum among all available pairs. It is simple to prove that if an algorithm starts with an empty matching and only adds closest available pairs to it until it is complete, the resulting matching is stable.

\subsection{Circle-Growing Algorithm}

In this section we describe our main practical algorithm, the \textit{circle-growing algorithm}, which mimics the continuous construction from~\cite{hoffman2006}.
First, we obtain the list of all the lattice points with coordinates ranging from $-n$ to $n$ sorted by distance to the origin. The resulting list $P$ emulates a circle growing from the origin. When initializing $P$, we can gain a factor of eight savings in space by sorting and storing only the points in the triangle $\triangle(0,0)(0,n)(n,n)$. The remaining points can be obtained by symmetry: if $p=(x,y)$ is a point in the triangle, the eight points with coordinates of the form $(\pm x,\pm y)$ and $(\pm y,\pm x)$ are at the same distance from the origin as $p$. Moreover, in applications where we find multiple stable grid matchings, such as in the stable $k$-means method, we  need only initialize $P$ once. The way we use $P$ depends on the type of centers we consider.

\ifdefined\ARXIV
\paragraph{Integer Centers (Algorithm~\ref{alg:cgi}).}
\else
\paragraph{Integer Centers.}
\fi
In this case we can use the fact that if we relocate the points in $P$ relative to a center, then they are in the order in which a circle growing from that center would reach them. To respect that all the circles grow at the same rate, we iterate through the points in $P$ in order. For each point $p$, we relocate it relative to each center $c$ to form the site $p+c$ (the order of the centers does not matter). We add to the matching any available center--site pair $(c,p+c)$. We iterate through $P$ until the matching is complete.

We require $O(n^2)$ space and $O(n^2\log n)$ time to sort the points in $P$. For the Euclidean metric instead of using distances to sort $P$ we can use squared distances, which take integer values between $0$ and $2n^2$. Then, we can use an integer sorting algorithm such as counting sort to sort in $O(n^2)$ time~\cite[Chapter~8.2]{Cormen2001}. Since each point in $P$ results in up to $O(k)$ center--site pairs, we need $O(n^2k)$ time to iterate through $P$.

\ifdefined\ARXIV
\begin{algorithm}
\caption{Circle growing algorithm for $k$ integer centers on an $n\times n$ grid.}
\label{alg:cgi}
\begin{algorithmic}
\State Set all sites as unmatched.
\State Set the quota of the first $n^2 \mbox{ mod } k$ centers to $\lceil n^2/k\rceil$.
\State Set the quota of the remaining centers to $\lfloor n^2/k\rfloor$.
\State Let $P=\mbox{list of points }(x,y)\mbox{ such that }-n<x,y<n$.
\State Sort $P$ by nondecreasing distance to $(0,0)$.
\ForAll{$p\in P$} until the matching is complete
        \ForAll{centers $c$ with quota $>0$}
            \State $s\gets p+c$
            \If{$0\leq s_x,s_y<n\mbox{ and }s\mbox{ is still available}$}
                \State Match $s$ and $c$.
                \State Reduce the quota of $c$  by 1.
            \EndIf
        \EndFor
\EndFor
\end{algorithmic}
\end{algorithm}
\else
\fi

\paragraph{Real Centers (Algorithm~\ref{alg:cgr}).}
If centers have real coordinates, we cannot translate the points in $P$ relative to the centers, because $p+c$ is not necessarily a lattice point.
The workaround is to associate each center $c$ to its closest lattice point $p_c$. Let $\delta$ be the maximum distance $d(c,p_c)$ among all centers. Then, the center--site pairs ``generated'' by each point $p$ in $P$ have the form $(c, p+p_c)$ and their distances can vary between $d(p,O)-\delta$ and $d(p,O)+\delta$ (where $O$ denotes the origin, $(0,0)$). Consequently, the distances of pairs generated by points $p_i,p_j$ in $P$ with $i<j$ may intertwine, but only if $d(p_j,O)-\delta\leq d(p_i,O)+\delta$. The points in $P$ after $p_i$ whose pairs might intertwine with those of $p_i$ form an annulus centered at $O$ with small radius $d(p_i,O)$ and big radius
\ifdefined\ARXIV
    $d(p_i,O)+2\delta$ (see Figure~\ref{fig:annulus}).
\else
 $d(p_i,O)+2\delta$.
\fi

Since $\delta$ is a constant (for the Euclidean metric, $\delta\leq \sqrt{2}/4$), it can be derived from the Gauss circle problem that such an annulus contains $O(d(p_i,O))=O(n)$ points.

\ifdefined\ARXIV
\begin{figure}[b!]
\centering
\includegraphics[width=0.48\linewidth]{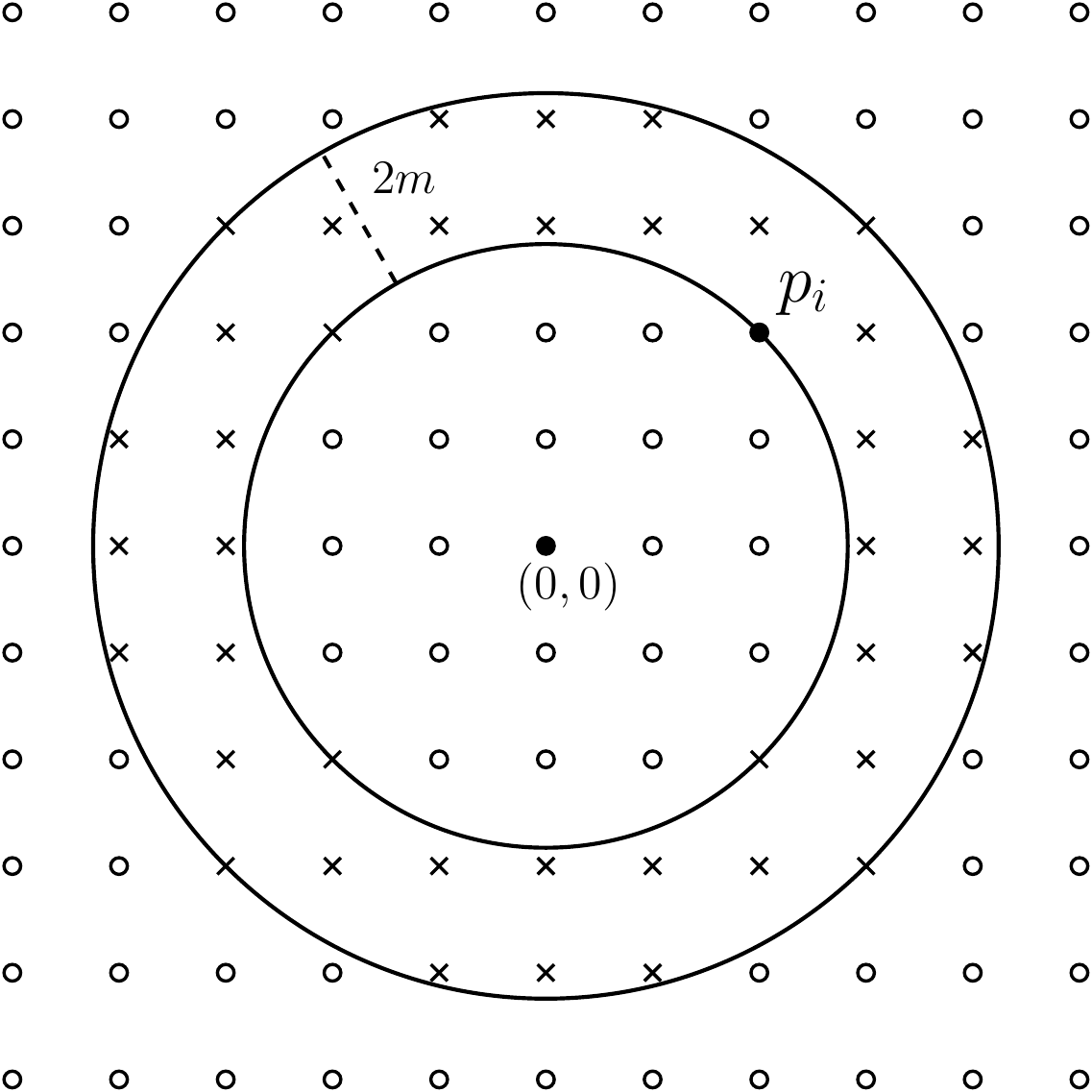}
\caption{The set of lattice points appearing after $p_i$ in $P$ whose pairs might intertwine with those of $p_i$ form an annulus centered at $O$ with small radius $d(p_i,O)$ and big radius $d(p_i,O)+2\delta$. In the figure, they are marked with an $\times$.}
\label{fig:annulus}\end{figure}
\else
\fi

The algorithm processes the points in $P$ in chunks of $n$ at a time, adding available center--site pairs generated by points in the chunk (or points after it, as we will see) to the matching in order by distance. The invariant is that after a chunk is processed, its points do not generate any more available pairs, and we can move on to the next one until the matching is complete. To do this, for each chunk we construct the list $L$ of all the pairs generated by its points. Let $d$ be the maximum distance among these pairs. If $p_i$ is the last point in the chunk, the points in $P$ from $p_{i+1}$ up to the last point at distance to the origin at most $d(p_{i},O)+2\delta$ can generate pairs with distance less than $d$. We add any such pair to $L$. We have to check $O(n)$ additional points, so $L$ still has size $O(kn)$. We sort all these pairs and consider them in order, adding any available pair to the matching.
Since each chunk has size $n$, there will be $O(n)$ chunks. Each one requires sorting a list of $O(kn)$ pairs, which requires $O(kn\log n)$ time (since $k\leq n^2$) and $O(kn)$ space. In total, we need $O(n^2k \log n)$ time and $O(n^2 + nk)$ space.

\begin{algorithm}
\caption{Circle growing algorithm for $k$ real centers on an $n\times n$ grid.}
\label{alg:cgr}
\begin{algorithmic}
\State Set all sites as unmatched.
\State Set the quota of the first $n^2 \mbox{ mod } k$ centers to $\lceil n^2/k\rceil$.
\State Set the quota of the remaining centers to $\lfloor n^2/k\rfloor$.
\State Let $P=\mbox{list of points }(x,y)\mbox{ such that }-n<x,y<n$.
\State Sort $P$ by nondecreasing distance to $(0,0)$.
\State For each center $c$, let $p_c=(\mbox{round}(c_x),\mbox{round}(c_y))$.
\State Let $\delta=\max\{\mbox{dist}(c,p_c)\}$ among all centers.
\State $j\gets 1$
\While{the matching is not complete}
    \State $L\gets\mbox{ empty list}$
    \State $i\gets \min(j+n, |P|)$
    \ForAll{$p\in P_j,\ldots,P_i$}\Comment{Add to $L$ pairs generated by points in the next chunk} 
            \ForAll{centers $c$ with quota $>0$}
                \State $s\gets p+p_c$
                \If{$0\leq s_x,s_y<n\mbox{ and }s\mbox{ is still available}$}
                    \State Add $(c,s)$ to $L$.
                \EndIf
            \EndFor
    \EndFor
    \State Let $d=\max\{\mbox{dist}(c,s)\}$ among all pairs $(c,s)\in L$.
    \ForAll{$p\in P_{i+1},\ldots,P_{|P|}$}
    \Comment{Add to $L$ pairs closer than pairs already in $L$}
        \If{$\mbox{dist}(p,O)>\mbox{dist}(P_i,O)+2\delta$}
            \State \textbf{break}
        \EndIf
            \ForAll{centers $c$ with quota $>0$}
                \State $s\gets p+p_c$
                \If{$0\leq s_x,s_y<n\mbox{ and }s\mbox{ is still available and dist}(c,s)\leq d$}
                    \State Add $(c,s)$ to $L$.
                \EndIf
            \EndFor

    \EndFor
    \State Sort $L$ by nondecreasing center--site distance.
    \ForAll{$(c,s)\in L$}
        \If{$c$ and $s$ are available}
            \State Match $s$ and $c$.
            \State Reduce the quota of $c$ by 1.
        \EndIf
    \EndFor
    \State $j\gets i+1$
\EndWhile
\end{algorithmic}
\end{algorithm}

\subsection{Distance-Sorting Methods}

Unless the centers are clustered together, the circle-growing algorithm finds many available pairs in the early iterations. However, it reaches a point in which most circles overlap. Even if the centers are randomly distributed, in the typical case a large fraction of centers have ``far outliers'', sites which belong to their region but are arbitrarily far because all the area in between is claimed by other centers. Consequently, many centers have to scan a large fraction of the square. At some point, thus, it is convenient to switch to a different algorithm that can find the closest available pairs quickly. In this section, let $m$ and $k\leq m$ denote, respectively, the number of available sites and centers after a matching has been partially completed.

\ifdefined\ARXIV
\paragraph{Pair Sort (Algorithm~\ref{alg:ps}).}
\else
\paragraph{Pair Sort Algorithm.}
\fi
This algorithm simply sorts all the center--site pairs by distance and considers them in order, adding any available pair to the matching until it is complete. This algorithm is convenient when we can use integer sorting techniques, as in the case of the Euclidean metric and integer centers. Then, it requires $O(mk)$ time and space. 

While the pair sort algorithm has a big memory requirement to be used starting with an empty matching, used after the circle-growing algorithm has matched a large fraction of sites results in improved performance.

\ifdefined\ARXIV
\begin{algorithm}
\caption{Pair Sort algorithm for $k$ centers and $m$ sites.}
\label{alg:ps}
\begin{algorithmic}
\State $L\gets\mbox{ empty list}$
\ForAll{centers $c$}
    \ForAll{sites $s$}
        \State Add $(c,s)$ to $L$.
    \EndFor
\EndFor
\State Sort $L$ by nondecreasing center--site distance.
\ForAll{$(c,s)\in L$}
    \If{$c$ and $s$ are available}
        \State Match $s$ and $c$.
        \State Reduce the quota of $c$ by 1.
    \EndIf
\EndFor
\end{algorithmic}
\end{algorithm}
\else
\fi

\ifdefined\ARXIV
\paragraph{Pair Heap (Algorithm~\ref{alg:ph}).}
\else
\paragraph{Pair Heap Algorithm.}
\fi
When centers have real coordinates, sorting all the pairs takes $O(mk\log m)$ time, but we can do better. We find for each site $s$ its closest center $c_s$, and build a min-heap with all the center--site pairs of the form $(c_s,s)$ using $d(c_s,s)$ as key. Clearly, the top of the heap is a closest available pair. We can iteratively extract and match the top of the heap until one of the centers becomes unavailable. When a center $c$ becomes unavailable, all the pairs in the heap containing $c$ become unavailable. At this point, there are two possibilities:
\setlist[description]{font=\normalfont\itshape}
\begin{description}
\item[Eager update]~
We find the new closest available center of all the sites that had $c$ as closest center and rebuild the heap from scratch so that it again contains one pair for each available site and its closest available center.
\item[Lazy update]~
We proceed as usual until we actually extract a pair $(c_s, s)$ with an unavailable center. Then, we find the new closest available center only for $s$, and reinsert the new pair in the heap. 
\end{description}
In both cases, we repeat the process until the matching is complete.

We have not addressed yet how to find the closest center to a site. For this, we can use a nearest neighbor (NN) data structure that supports deletions. Such a data structure maintains a set of points and is able to answer \textit{nearest neighbor queries}, which provide a query point $q$ and ask for the point in the set closest to $q$. For the pair heap algorithm, we initialize the NN data structure with the set of centers and delete them as they become unavailable.

Since we need deletions we can use a \textit{dynamic} NN data structure, i.e., with support for insertions as well as deletions. The simplest NN algorithm is a linear search, and a dynamic data structure based on it has $O(k)$ time per query and $O(1)$ time per update. The best known complexity of a dynamic NN data structure is $O(\log^5 k)$ amortized time per operation~\cite{Chan2010,KapMulRod-16}.

\ifdefined\ARXIV
\begin{algorithm}
\caption{Pair Heap algorithm with lazy updates for $k$ centers and $m$ sites.}
\label{alg:ph}
\begin{algorithmic}
\State Let $C=\mbox{nearest neighbor data structure with all the centers.}$
\State Let $H=\mbox{empty min-heap of center--site pairs using distance as key.}$
\ForAll{sites $s$}
    \State Add $(C.\mbox{nearest}(s), s)$ to $H$.
\EndFor
\While{$H$ is not empty}
    \State $(c,s)\gets H.\mbox{removeMin}()$
    \If{$c$ has quota $>0$}
        \State Match $s$ and $c$.
        \State Reduce the quota of $c$ by 1.
    \Else
        \State Remove $c$ from $C$.
        \State Add $(C.\mbox{nearest}(s), s)$ to $H$.
    \EndIf
\EndWhile
\end{algorithmic}
\end{algorithm}
\else
\fi

Given that we know all the query points for our NN data structure ahead of time (the sites), we can build for each site $s$ an array $A_s$ with all the centers sorted by distance to $s$. Then, the closest center to a site $s$ is $A_s[i_s]$, where $i_s$ is the index of the first available center in $A_s$.
When a center is deleted we simply mark it. When we get a query for the closest center to a site $s$, we search $A_s$ until we find an unmarked center. We can start the search from the index of the center returned in the last query for $s$.
This data structure requires $O(mk)$ space and has a $O(mk\log k)$ initialization cost to sort all the arrays.
The interesting property is that if we do $O(k)$ queries for a given site $s$, we require $O(k)$ time for all of them, as in total we traverse $A_s$ only once.
We call this data structure \textit{presort}, although it is not strictly a NN data structure because it knows the query points ahead of time.

In the pair heap algorithm, we can combine eager and lazy updates with any NN data structure. In any case, the running time is influenced by $\alpha$, the sum among all centers $c$ of the number of sites that had $c$ as closest center when $c$ became unavailable. In the worst case $\alpha=O(km)$, but assuming that each center is equally likely to be the closest center to each site, the expected value of $\alpha$ is $O(m)$. 
\ifdefined\ARXIV
In Appendix~\ref{app:alphaexperiment} we test the value of $\alpha$ empirically.
\else
In the full version of the paper we test the value of $\alpha$ empirically in several different settings, and in every case we find $\alpha<10m$.
\fi

With eager updates in total we have to initialize the NN data structure, perform $m$ \textit{extract-min} operations, $O(m+\alpha)$ NN queries, $k$ NN deletions, and rebuild the heap $k$ times. Thus, the running time is $O(P(k,m)+m\log m+(m+\alpha) Q(k)+kD(k)+km)$, where $P(k,m)$ is the cost  of initializing the NN data structure of choice with $k$ points (and $m$ query points, in the case of the presort data structure), and $Q(k)$ and $D(k)$ are the costs of queries and deletions, respectively. With lazy updates, instead of rebuilding the heap we have $O(\alpha)$ extra \textit{insert} and \textit{extract-min} heap operations, which requires $O(\alpha\log m)$ time.

For real centers, the best worst-case bound is with eager deletions and the presort NN data structure. In that case, we have that the NN queries take $O(km)$ for any $\alpha$, so the total running time is $O(mk\log k+m\log m)$. If we assume that $\alpha=O(m)$, then the best time is with lazy deletions and the NN data structure from~\cite{Chan2010,KapMulRod-16}.
The running time with this heuristic assumption is $O(m\log^5 k+m\log m)$.

\subsection{Bichromatic Closest Pairs and Nearest Neighbor Chains}

We now describe a less-practical solution based on bichromatic closest pairs which achieves the best theoretical running time that we have been able to prove.
A bichromatic closest pair (BCP) data structure maintains a set of points, each colored red or blue, and is able to answer queries asking for the closest pair of different color.

The stable grid matching problem can be solved with a BCP data structure that supports deletions, either on its own or after the circle-growing algorithm.
We first initialize the data structure with the available sites and centers as blue and red points, respectively. Then, we repeatedly find and match the closest pair, remove the site, and remove the center if it becomes unavailable. The running time is $O(P(m)+mQ(m)+mD(m))$, where $P(m),Q(m),$ and $D(m)$ are the initialization, query, and deletion costs, respectively, for the BCP data structure of choice containing $m$ blue points and $k\leq m$ red points.

Eppstein~\cite{Eppstein1995} proposed a fully dynamic BCP data structure that uses an auxiliary dynamic NN data structure. Using it, the sequence of operations required to solve the stable grid matching problem takes $O(mT(m)\log^2 m)$ time, where $T(m)$ is the cost per operation of the NN data structure. In particular, combining this with the dynamic nearest neighbor data structure of Chan~\cite{Chan2010} and Kaplan et~al.~\cite{KapMulRod-16} gives a total time bound of $O(n^2\log^7 n)$ for this problem.

To improve this, we observe that (with a suitable tie-breaking rule to ensure that no two distances are equal) it is not necessary to find the bichromatic closest pair in each step: it suffices, instead, to find a mutual nearest neighbor pair: a pixel and a center that are closer to each other than to any other pixel or center. The reason is twofold. First, in the algorithm that repeatedly finds and removes closest pairs, every pair $(c,p)$ of mutual nearest neighbors eventually becomes a closest pair, because until they do, nothing else that the algorithm does can change the fact that they are mutual nearest neighbors. So $(c,p)$ will eventually become matched by the algorithm. Second, if we find a pair $(c,p)$ that will eventually become matched (such as a mutual nearest neighbor pair), it is safe to match them early; doing so cannot affect the correctness of the rest of the algorithm.

To find these, we may adapt the \emph{nearest-neighbor chain algorithm} from the theory of hierarchical clustering~\cite{Ben-CAD-82,Jua-CAD-82} which uses a stack to repeatedly find pairs of mutual nearest neighbors at a cost of $O(1)$ nearest neighbor queries per pair. In more detail, the algorithm is as follows.

\begin{enumerate}
\item Initialize two dynamic nearest neighbor structures for the pixels and centers, and an empty stack $S$.
\item Repeat the following steps until all pixels have been matched:
\begin{enumerate}
\item If $S$ is empty, push an arbitrary point (either a pixel or a center) onto $S$.
\item Let $p$ be the point at the top of $S$, and use the nearest neighbor data structure to find the nearest point $q$ of the opposite color to $p$.
\item If $q$ is not already on $S$, push it onto $S$. Otherwise, $q$ must be the second-from-top point on $S$, and is a mutual nearest neighbor with $p$. Pop $p$ and $q$, match them to each other, and remove one or both of $p$ or $q$ from the nearest neighbor data structure (always remove the pixel, and remove the center if it becomes unavailable).
\end{enumerate}
\end{enumerate}

Note that in step 2.\,(c) $q$ must be second-from-top because we have a cycle of (non-mutual) nearest neighbors starting with $p\rightarrow q$ and then up the stack back to $p$. At each step along this cycle, the distance decreases or stays equal. But it cannot decrease, because there would be no way to increase back again, and nothing but $q\rightarrow p$ can be equal to $p\rightarrow q$, because we are using a tie-breaking rule. So the cycle has length two and $q$ is second-from-top.

Each step that pushes a new point onto $S$ can be charged against a later pop operation and its associated matched pixel, so the number of repetitions is $O(n^2)$.
This algorithm gives us the following theorem.

\begin{theorem}
The stable grid matching problem can be solved in $O(n^2)$ operations of a dynamic nearest neighbor data structure. In particular, with the structure of Chan~\cite{Chan2010} and Kaplan et~al.~\cite{KapMulRod-16}, the time is $O(n^2\log^5 n)$.
\end{theorem}

\section{Experiments}

\paragraph{Datasets.}
Table~\ref{tab:dataset} summarizes the parameters used in the different experiments. We use the following labels for the algorithms: $CG$ the circle-growing algorithm alone, and $PS$ and $PH$ for the combination of CG and the pair sort and pair heap algorithms, respectively. Moreover, for the pair heap algorithm we consider the following variations: eager/presort ($PH_{E,P}$), eager/linear search ($PH_{E,L}$), lazy/presort ($PH_{L,P}$), and lazy/linear search ($PH_{L,L}$).

\ifdefined\ARXIV
We focus on the Euclidean metric, but Appendix~\ref{app:moreplots} has all the same figures for the Manhattan and Chebyshev metrics.
\else
We focus on the Euclidean metric, but in the full version of the paper we also consider the Manhattan and Chebyshev metrics.
\fi
The parameter $n$ is the length of  the side of the square grid, and $k$ is the number of centers. In all the experiments, the centers are chosen uniformly and independently at random. Moreover, every data point is the average of 10 runs, each starting with different centers.

The cutoff is the parameter used to determine when to switch from the circle-growing algorithm to a different one. We define it as a ratio between the number of available pairs and the number of pairs already considered by the circle-growing algorithm.

The algorithms were implemented in C++ (gcc version 4.8.2) and the interface in Qt. The experiments were executed by a Intel(R) Core(TM) CPU i7-3537U 2.00GHz with 4GB of RAM, on Windows 10.
\begin{table}[b]
\centering
\caption{Summary of parameters used in the experiments section.}
\label{tab:dataset}
\begin{tabular}{|l|l|l|r|r|r|}
\hline
\textbf{Experiment}          & \textbf{Algorithms}      & \textbf{Metric} & \multicolumn{1}{l|}{\textbf{$n$}} & \multicolumn{1}{l|}{\textbf{$k$}} & \multicolumn{1}{l|}{\textbf{Cutoff}} \\ \hline
\textbf{Exec. time (Fig.~\ref{fig:exectimeL2})} & All                      & $L_2$          & varies                            & $10n$                             & 0.15                                 \\ \hline
\textbf{Cutoff (Fig.~\ref{fig:cutoffL2})}     & $CG,PH_{L,L}$            & $L_2$           & 1000                              & varies                            & varies                               \\ \hline
\ifdefined\ARXIV
\textbf{Cutoff (Fig.~\ref{fig:profileL2})}     & $CG,PH_{L,L}$            & $L_2$           & 1000                              & 10000                             & varies                               \\ \hline
\textbf{Value of $\alpha$ (Appendix~\ref{app:alphaexperiment})}   & $PH$ with lazy deletions & $L_2$           & 1000                              & varies                            & ---                                  \\ \hline
\else
\fi
\end{tabular}
\end{table}

\begin{figure}
\centering
\ifdefined\ARXIV
\includegraphics[width=0.99\linewidth]{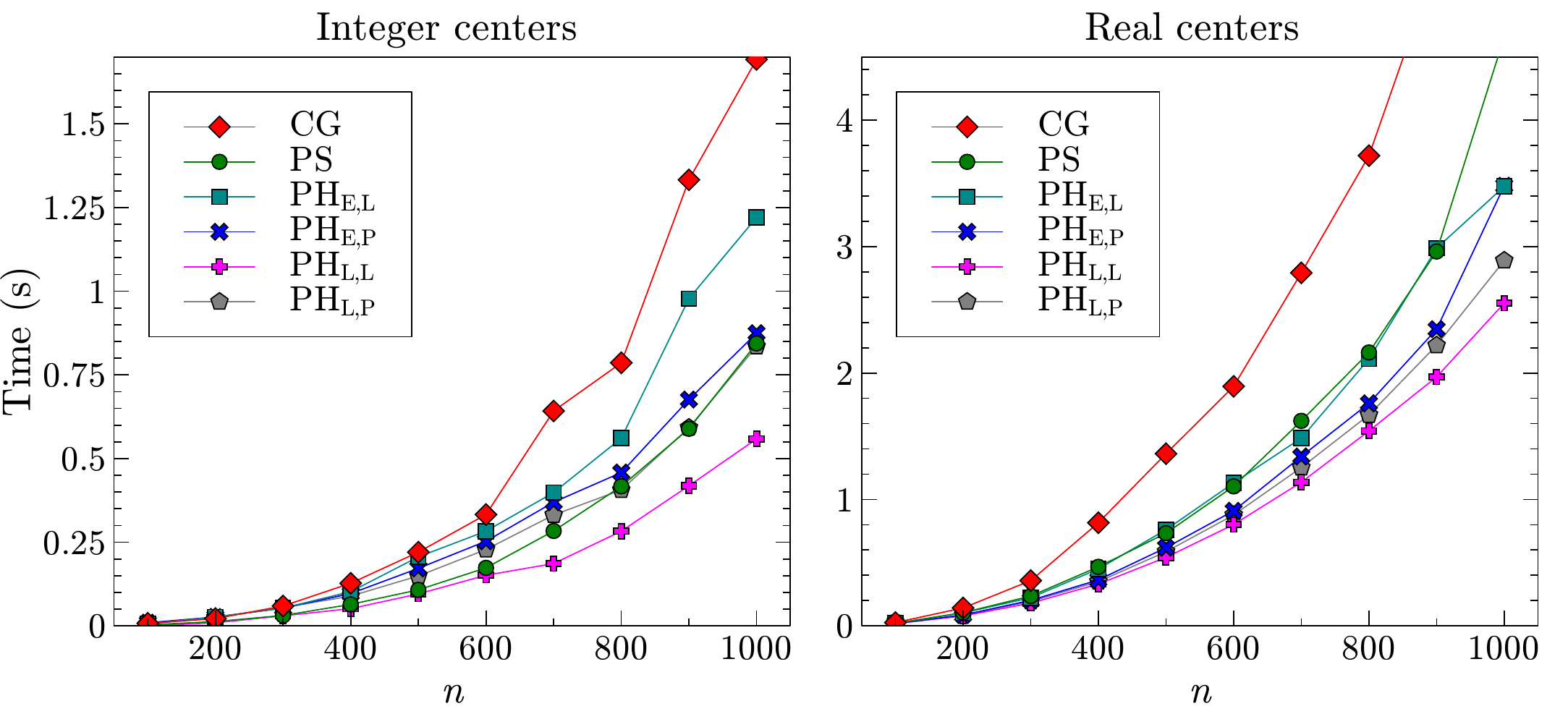}
\else
\includegraphics[width=0.8\linewidth]{figure11}
\fi
\caption{Execution time of the various algorithms for integer (left) and real (right) centers. For all the methods but \textit{CG}, the cutoff is $0.15$. Each data point is the average of 10 runs with $10n$ randomly distributed centers and the $L_2$ metric.}
\label{fig:exectimeL2}
\end{figure}

\paragraph{Algorithm Comparison.} Figure~\ref{fig:exectimeL2} contains a comparison of all the algorithms.
Pair heap is generally better than pair sort, even for integer distances where it has a higher theoretical complexity. Among pair heap variations, lazy/linear is the best for both types of centers. In general lazy updates perform better, but eager/presort is also a strong combination because they synergize: eager updates require more NN queries in exchange for less extract-min heap operations, and the presort data structure has fast NN queries.

\ifdefined\ARXIV
\else
\vspace{-0.1in}
\fi

\paragraph{Optimal Cutoff.} When combining the circle-growing algorithm with another algorithm, the efficiency of the combination depends on the cutoff used to switch between both. If we switch too soon, we don't exploit the good behavior of the circle-growing algorithm when circles are still mostly disjoint. If we switch too late, the circle-growing algorithm slows down as it grows the circles in every direction just to reach some outlying region.

\begin{figure}[t]
\centering
\ifdefined\ARXIV
\includegraphics[width=0.99\linewidth]{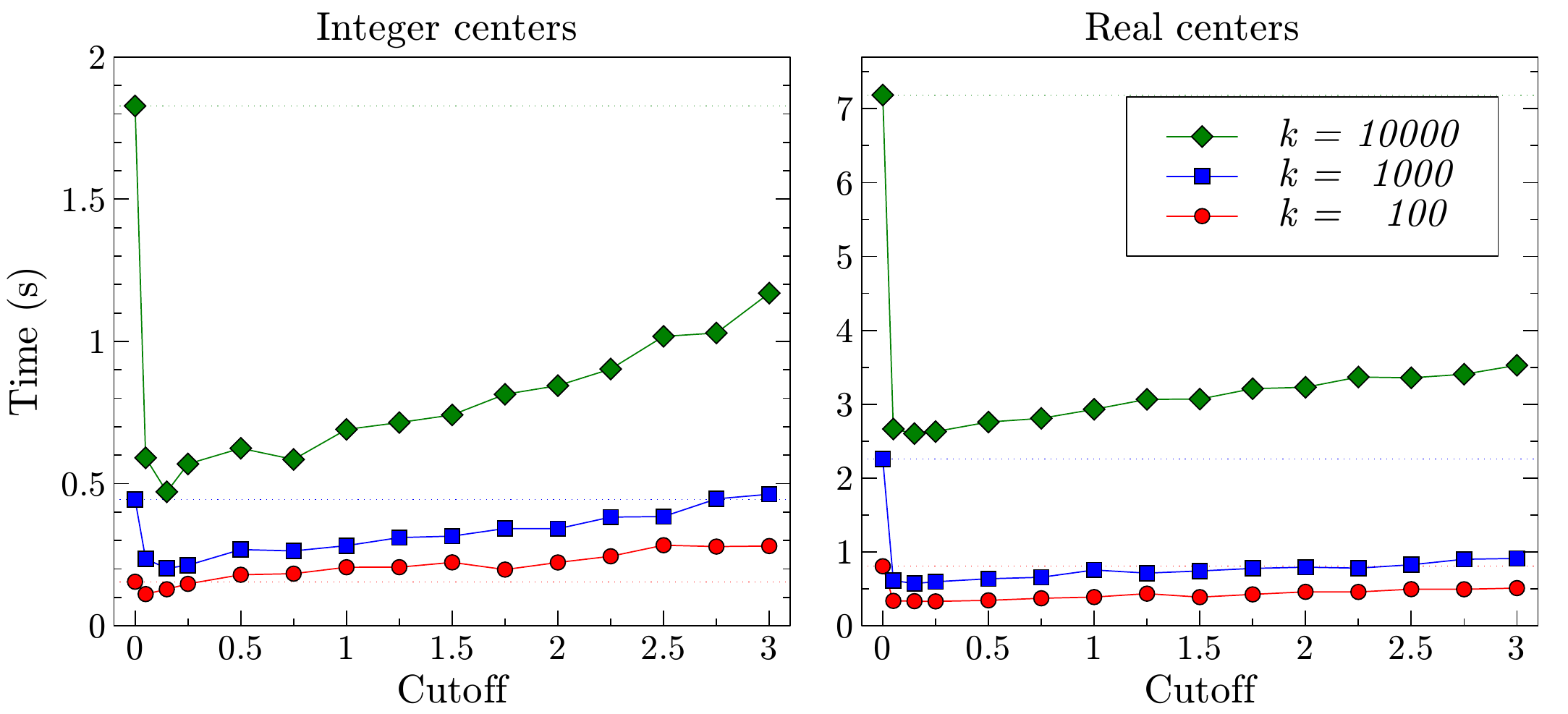}
\else
\includegraphics[width=0.8\linewidth]{figure9}
\fi
\caption{Execution time of the circle-growing algorithm for integer (left) and real (right) centers, combined with the pair heap algorithm with lazy updates and a linear search NN data structure. The dotted lines denote the running time of the circle-growing algorithm alone, i.e., with cutoff 0. Each data point is the average of 10 runs with randomly distributed centers, $n=1000$, and the $L_2$ metric.}
\label{fig:cutoffL2}
\end{figure}

Figure~\ref{fig:cutoffL2} illustrates the role of the cutoff.
It shows that most of the execution time of the circle-growing algorithm is spent with the very few last available pairs, so even a really small cutoff prompts a substantial improvement. After that, the additional time spent in the pair heap algorithm slightly beats the savings in the circle-growing algorithm, resulting in a steady increase of the total running time.
\ifdefined\ARXIV
In addition, Figure~\ref{fig:profileL2} shows in more detail how this execution time is divided among the circle-growing algorithm and the pair heap algorithm.
\else
\fi

\ifdefined\ARXIV
\begin{figure}[t]
\centering
\ifdefined\ARXIV
\includegraphics[width=0.99\linewidth]{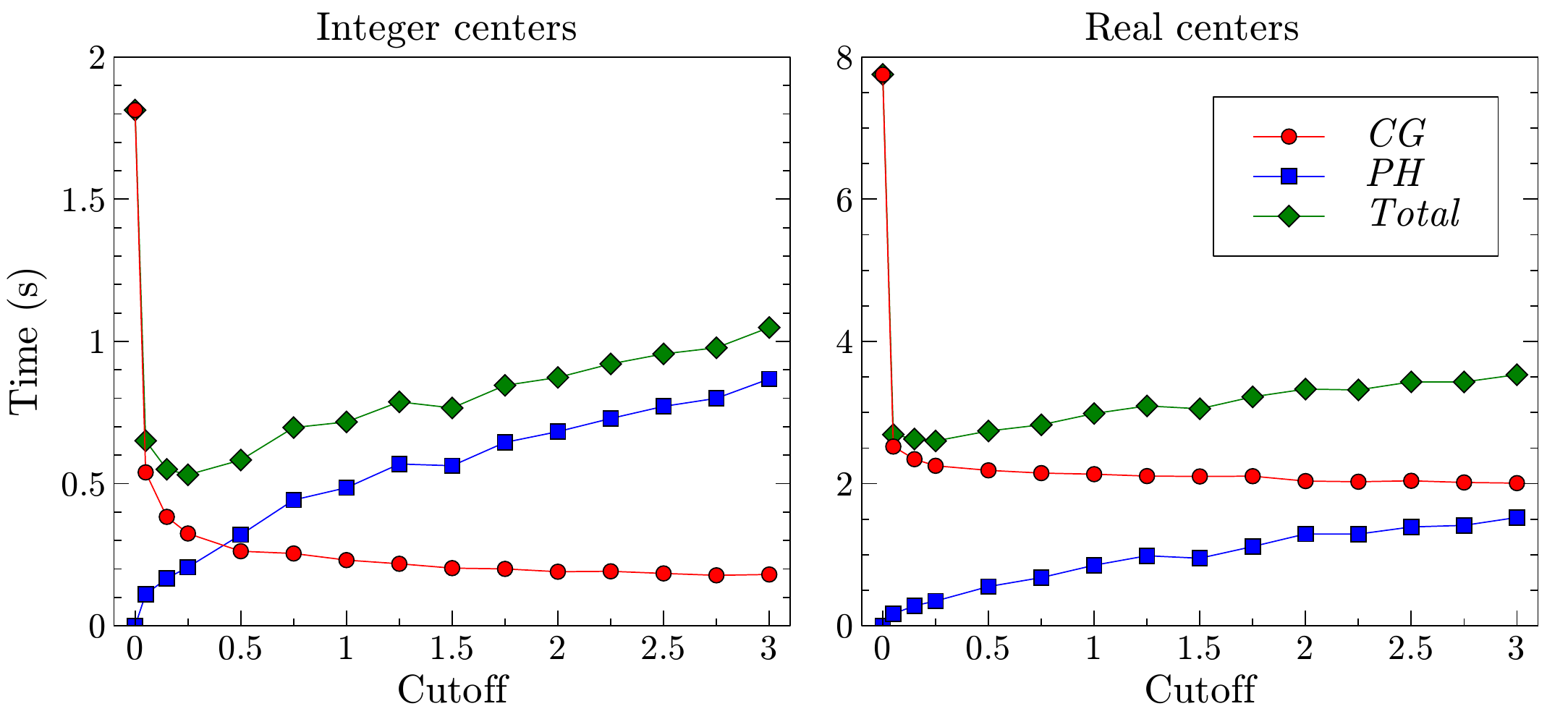}
\else
\includegraphics[width=0.8\linewidth]{figure10}
\fi
\caption{Execution time of the circle-growing algorithm for integer (left) and real (right) centers, combined with the pair heap algorithm with lazy updates and a linear search NN data structure. In addition to the total execution time, we show the execution time spent in each algorithm. Each data point is the average of 10 runs with $n=1000$, $k=10000$ randomly distributed centers, and the $L_2$ metric.}
\label{fig:profileL2}
\end{figure}
\else
\fi

\section{Discussion}

We have defined the stable grid matching problem, developed efficient theoretical algorithms and practical implementations of slower but simpler algorithms for this problem, and used our implementation to test different strategies for center placement in $k$-means like stable clustering algorithms. However, this work leaves several  open questions:
\begin{itemize}
\item For which $n$ and $k$ does the stable grid matching problem have a placement of centers for which all clusters are connected, and how can such centers be found?
\item Can the worst-case running time of our theoretical $O(n^2\log^5 n)$-time algorithm be improved? Is it possible to achieve similar runtimes without going through fully-dynamic bichromatic closest pair data structures?
\item Can we obtain practical algorithms whose runtime has lower worst-case dependence on $k$ than our $O(n^2k)$-time circle-growing and distance-sorting methods?
\item Our bichromatic closest pair and distance-sorting algorithms can be made to work for arbitrary point sets (not just pixels) but the circle-growing method assumes that the points form a grid, and its time analysis depends on the fact that the grid is a fat polygon (so that the area of each circle is proportional to the number of grid points that it covers) and that testing whether a point belongs to the grid is trivial. Can this method be extended to pixelated versions of more complicated polygons?
\item How efficiently can we perform similar distance-based stable matching problems for graph shortest path distances instead of geometric distances? Can additional structure (such as the structures found in real-world road networks) help speed up this computation?
\end{itemize}

\bibliographystyle{splncs03}
\bibliography{paper12}

\ifdefined\ARXIV
\else
\end{document}
\fi

\newpage

\appendix

\section{Stable $k$-means method with weighted centroids}\label{app:centroids}

\begin{figure}
\centering
\includegraphics[width=0.99\linewidth]{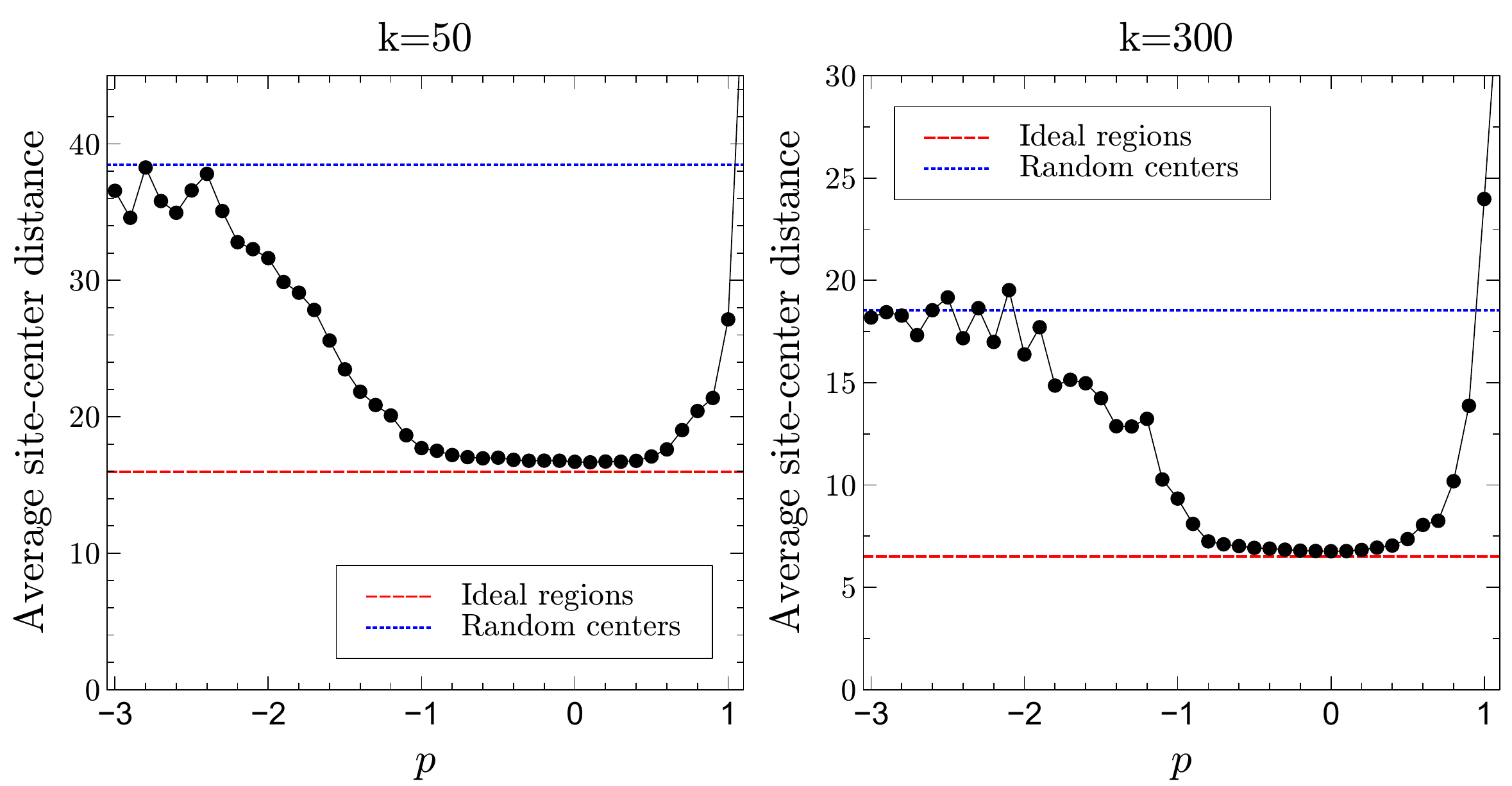}
\caption{Average distance among sites and their matched center after 100 iterations of stable $k$-means for different exponents $p$ of the weighted centroid. We consider integer centers, the Euclidean metric, and grid size $n=300$. The blue dotted line denotes the average distance with random centers, and the red dashed line denotes the average distance in an ideal region (i.e., a disk).  Each data point is the average of 10 runs starting with randomly distributed centers.}
\label{fig:weightedcentroid}
\end{figure}

\begin{figure}
\centering
\begin{minipage}[b]{0.42\linewidth}
\includegraphics[width=0.99\linewidth]{figure2}\\[4pt]
\includegraphics[width=0.99\linewidth]{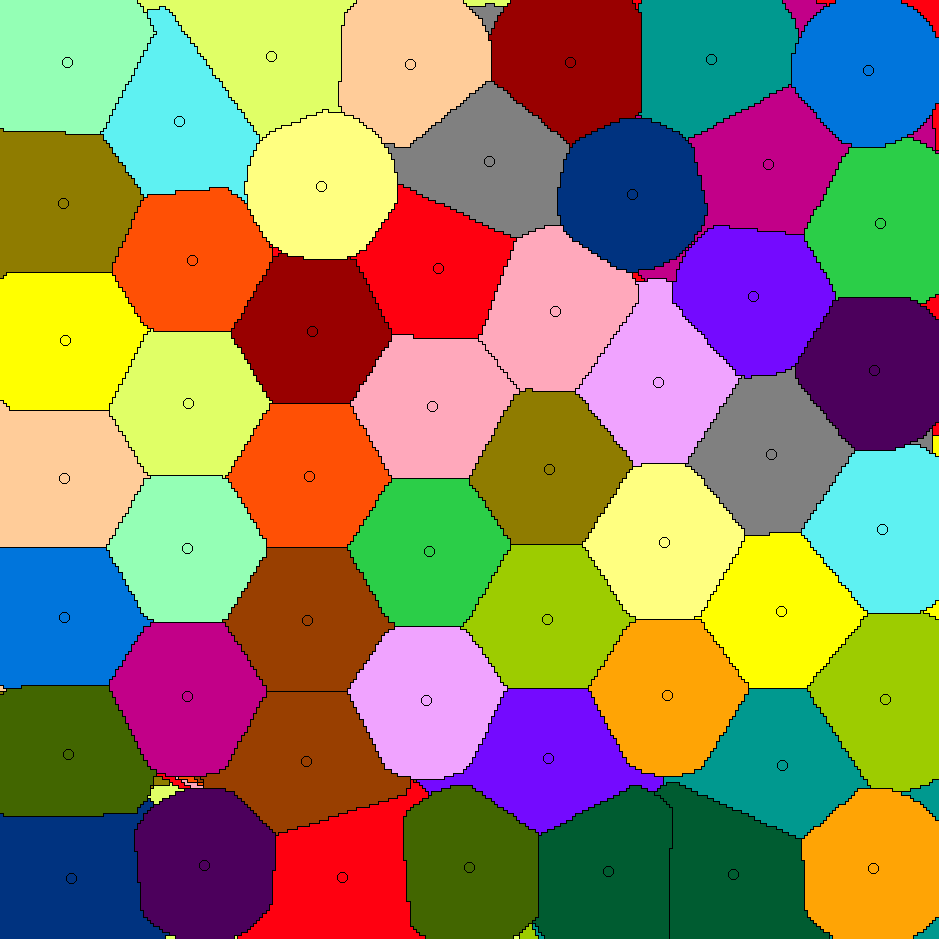}\\[4pt]
\includegraphics[width=0.99\linewidth]{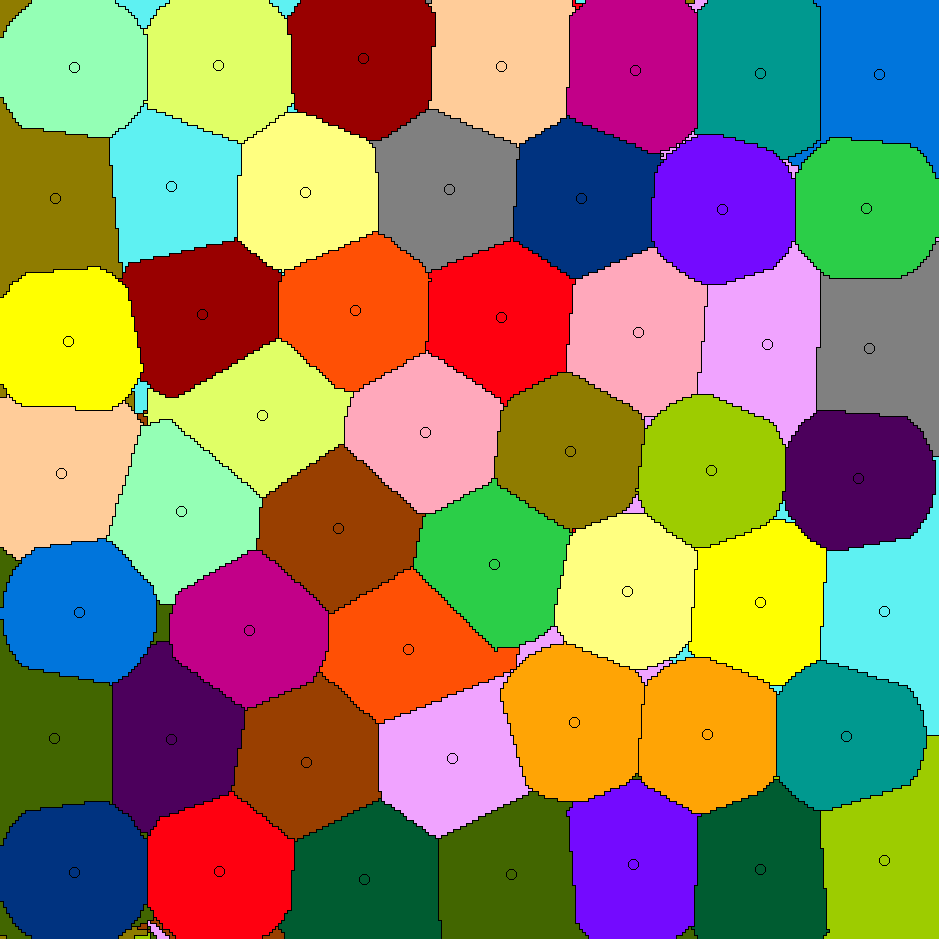}
\end{minipage}
\begin{minipage}[b]{0.42\linewidth}
\includegraphics[width=0.99\linewidth]{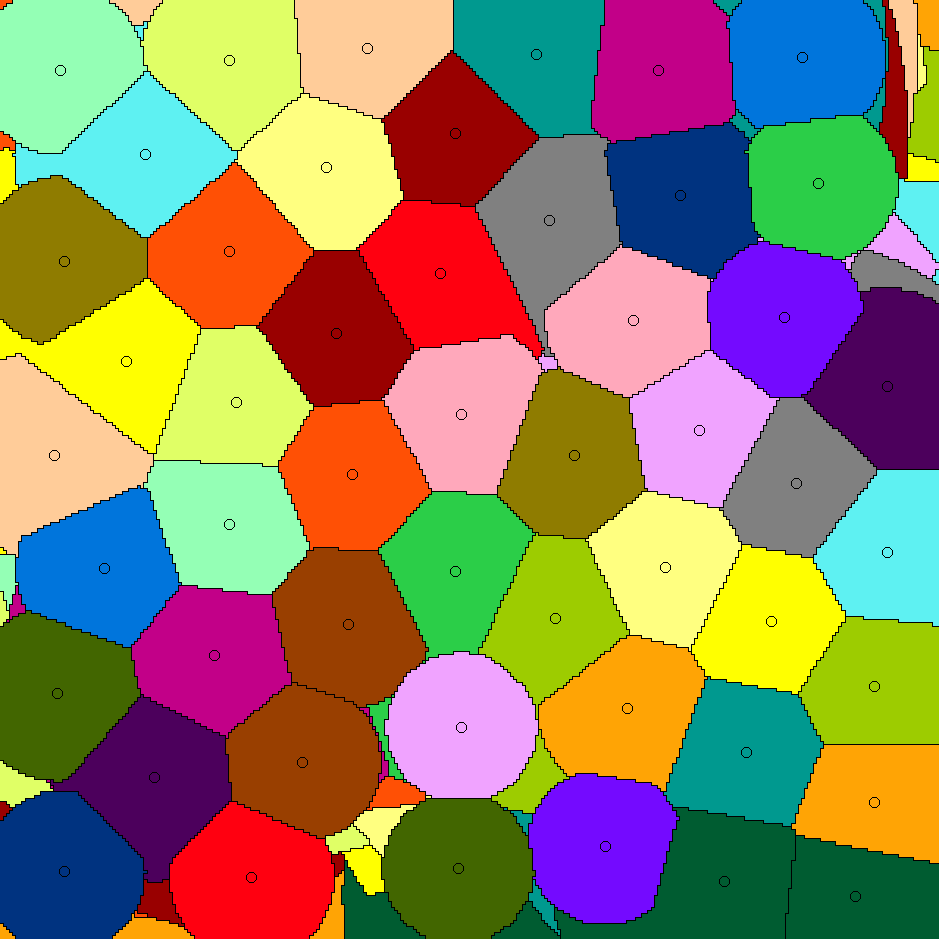}\\[4pt]
\includegraphics[width=0.99\linewidth]{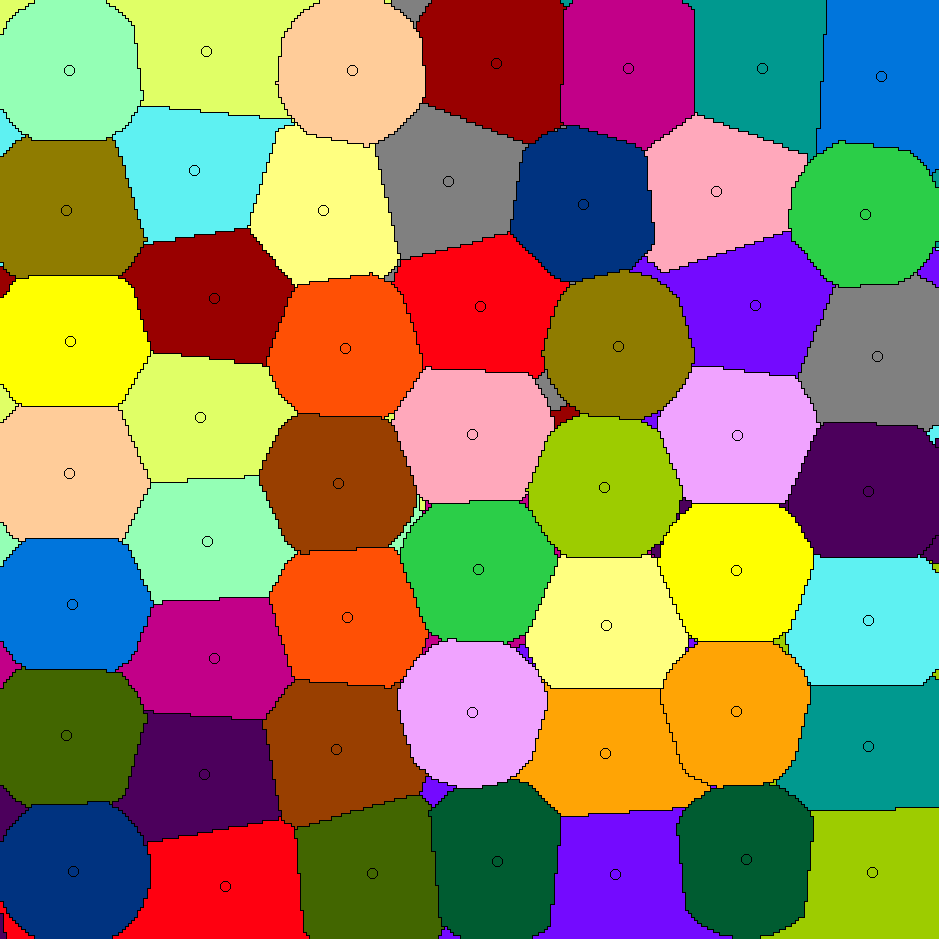}\\[4pt]
\includegraphics[width=0.99\linewidth]{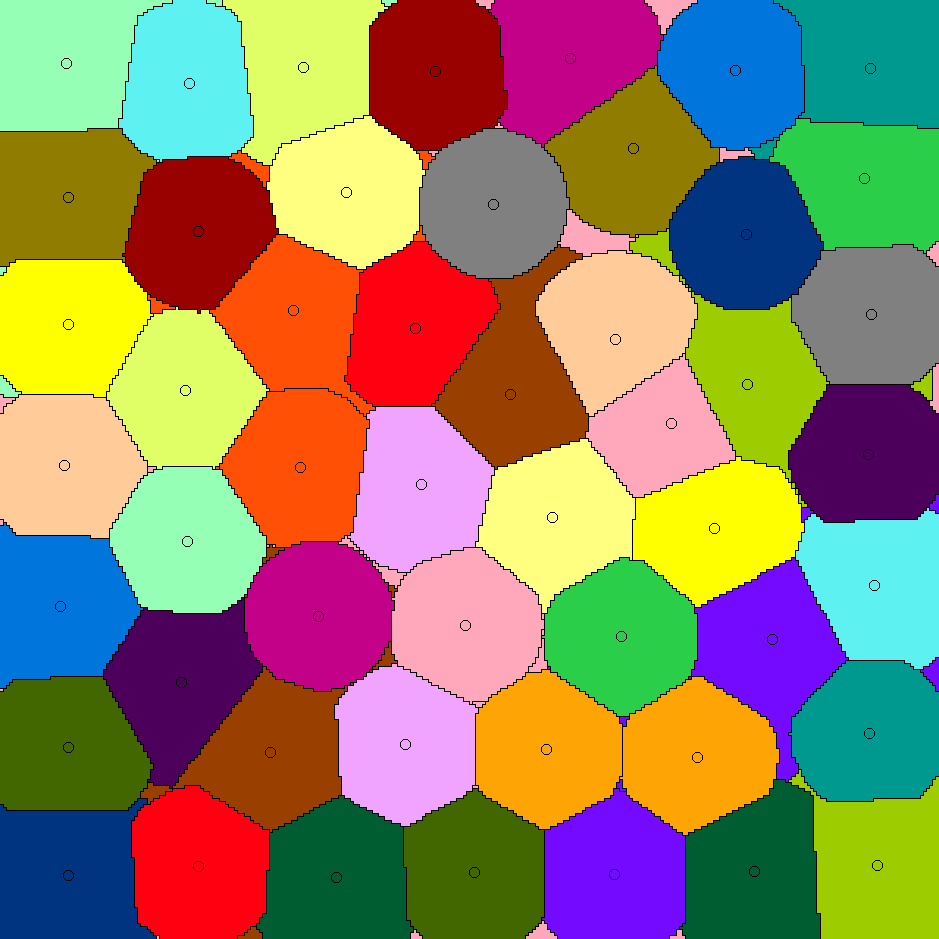}
\end{minipage}
\caption{Top left: Same matching from Figure~\ref{fig:kmeans} with $n=300$ and $50$ random centers. Other figures, from top to bottom and left to right: Result of the stable k-means method with weighted centroids for $p=-2,-1,0.1,0.2,$ and $0.5$.}
\label{fig:kmeans2}
\end{figure}

\begin{figure}
\centering
\begin{minipage}[b]{0.42\linewidth}
\includegraphics[width=0.99\linewidth]{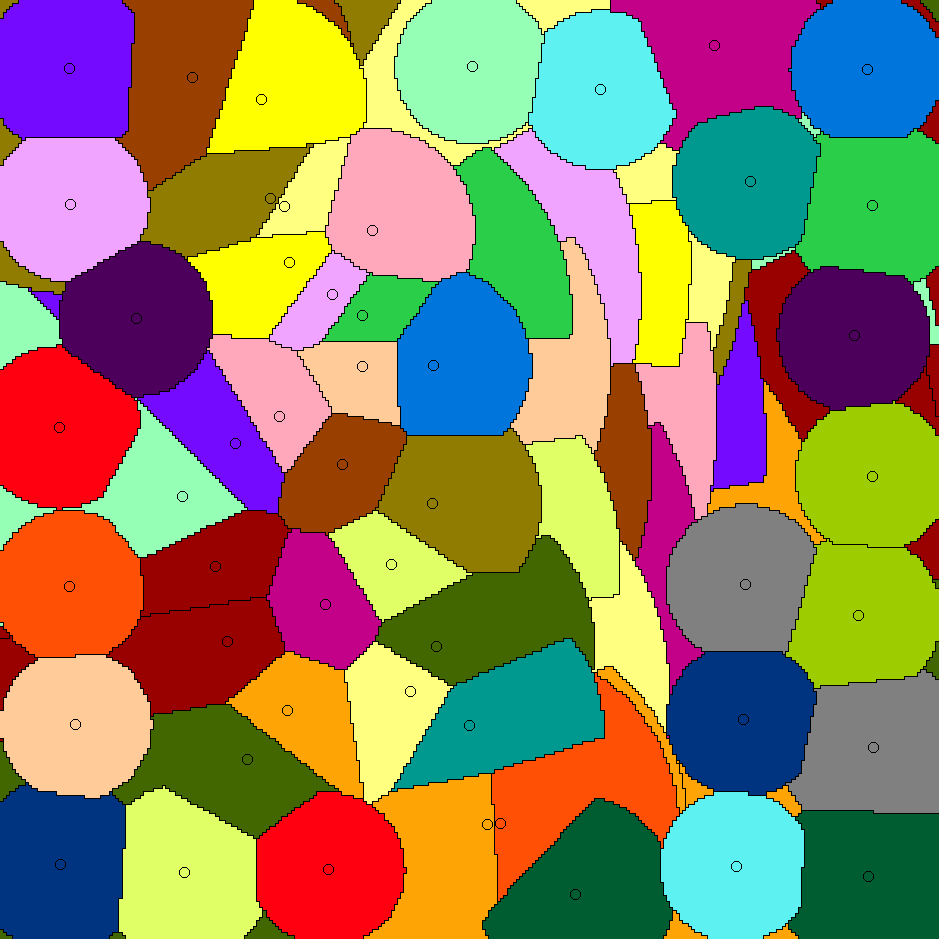}\\[4pt]
\includegraphics[width=0.99\linewidth]{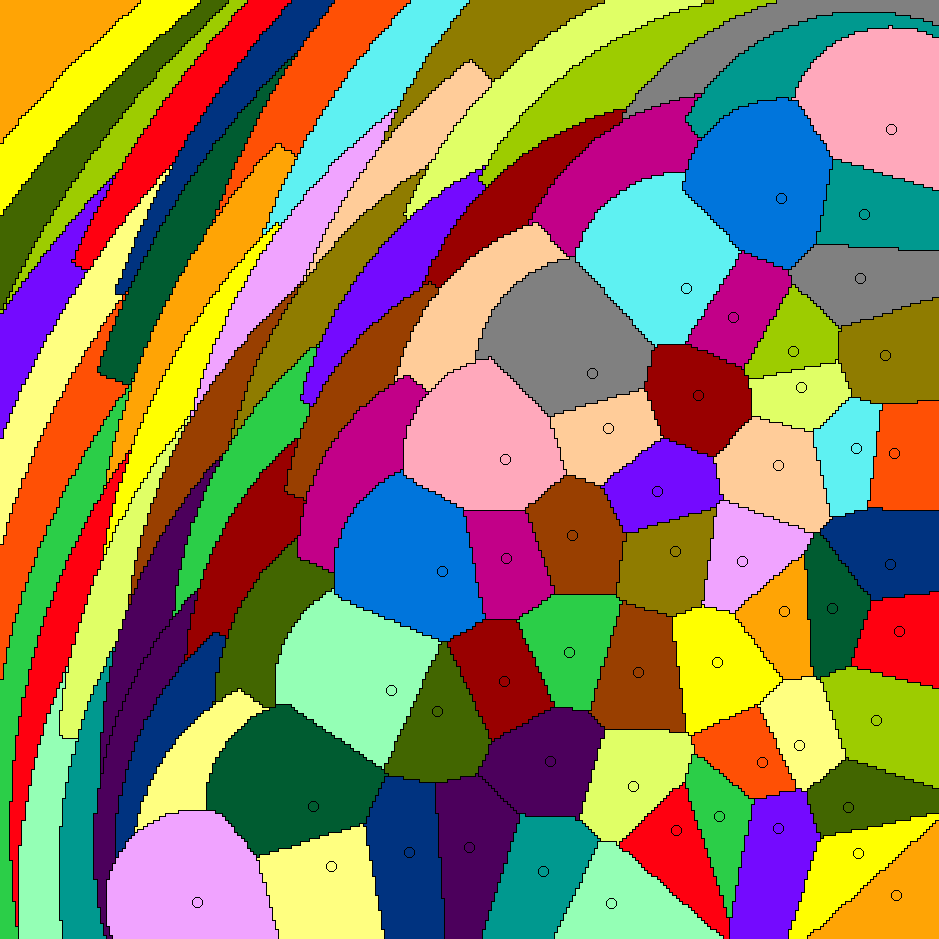}\\[4pt]
\end{minipage}
\begin{minipage}[b]{0.42\linewidth}
\includegraphics[width=0.99\linewidth]{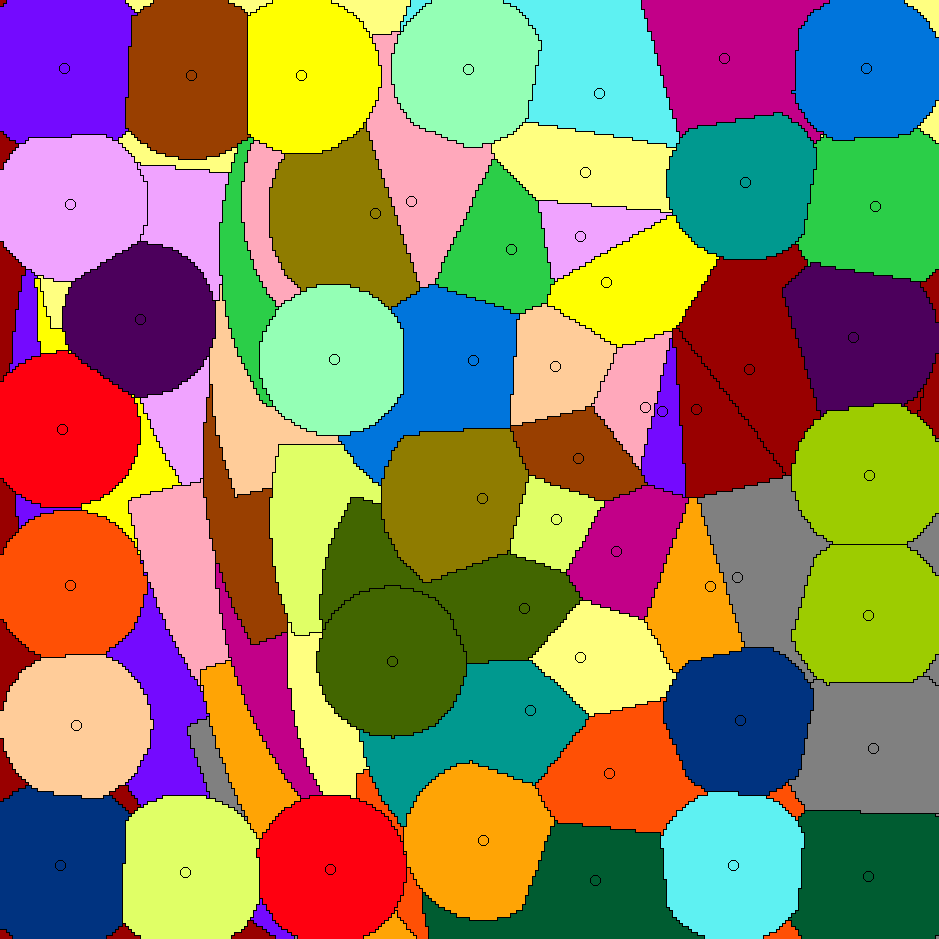}\\[4pt]
\includegraphics[width=0.99\linewidth]{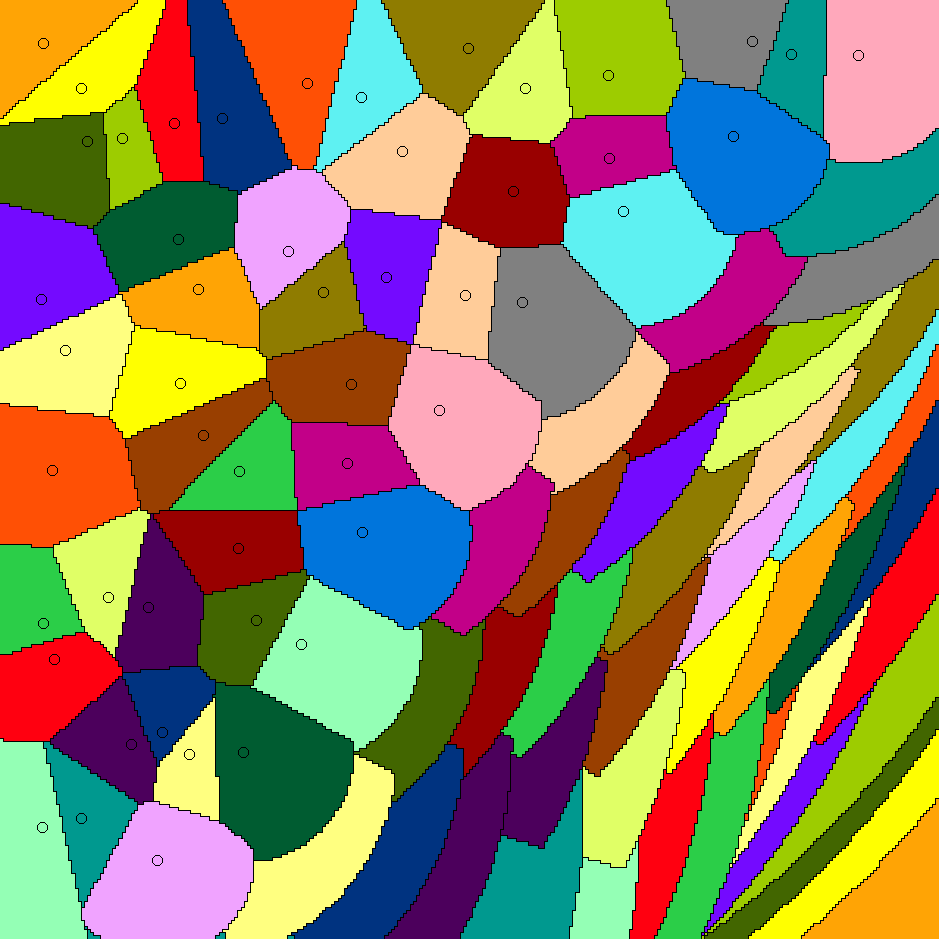}\\[4pt]
\end{minipage}
\caption{Top: Two consecutive iterations of the stable k-means method with weighted centroids for $p=1$, for the same matching from Figure~\ref{fig:kmeans} with $n=300$ and $50$ random centers. Bottom: same for $p=10$.}
\label{fig:kmeans3}
\end{figure}

Figure~\ref{fig:weightedcentroid} shows how the exponent of the weighted centroid affect the result of the stable $k$-means method. As evaluation measure, we use the average distance among sites and their matched center. The best results are with $-0.8<p<0.4$ (for different metrics and grid sizes, we obtain similar results).
Figure~\ref{fig:kmeans2} shows the result of the stable $k$-means method for the stable grid matching problem for values of $p$ (the exponent in the formula for the weighted centroids) between $-2$ and $0.5$. Figure~\ref{fig:kmeans3} shows the transient behavior of stable $k$-means with $p\geq 1$.

On a related note, if we set $p=-1$ and repeatedly move a center to its weighted centroid (keeping its region unchanged), we get Weiszfeld's algorithm, a known iterative method for finding the geometric median (the point minimizing the sum of distances to its region)~\cite{Weiszfeld2009}. In our case, we are not doing the same thing, because we are also recomputing the regions after each update of the centers, but it is still the case that for $p=-1$ the algorithm converges to a state where each center is the geometric median of its region.

\section{Value of $\mathbf{\alpha}$.}\label{app:alphaexperiment}
The running time of the pair heap algorithm depends on $\alpha$, the sum among all centers $c$ of the number of sites that had $c$ as closest center when $c$ became unavailable.
There is a gap between the worst case $\alpha=O(km)$ and the expected case $\alpha=O(m)$ when sites and centers are distributed randomly.
Even with randomly located centers, the distribution of remaining sites and centers after the circle-growing algorithm is not random, so here we are interested in the actual value of $\alpha$ in such cases. More precisely, we are interested in $\beta$, the total number of extra \textit{extract-min} operations (i.e., operations returning an unavailable pair) when using the pair heap algorithm with lazy updates. Note that $\beta\leq\alpha$, because with lazy updates when a center becomes unavailable some of the sites that would have it as closest center still have a previously unavailable center instead.

First, we observe the value of $\beta/m$ when using the pair heap algorithm on its own in an $n=100$ ($m=10000$) grid, using randomly distributed integer centers and the $L_2$ metric. The maximum values of $\beta/m$ among 10 runs for each $k$ were $0.64$ for $k=10$, $0.80$ for $k=100$, and $0.82$ for $k=1000$. We obtained similar values for the $L_1$ and $L_\infty$ metrics; in every case, $\beta<m$.

Second, we observed the values of $\beta/m$ when using the pair heap algorithm after the circle-growing algorithm, again using randomly distributed integer centers and the $L_2$ metric. We switched between algorithms when there were $m=10000$ available pairs. The maximum values of $\beta/m$ among 10 runs for each $k$ were $1.20$ for $k=100$ (with $4$ remaining centers), $4.42$ for $k=1000$ (with $31$ remaining centers), and $7.86$ for $k=10000$ (with $275$ remaining centers). We also obtained similar values for the $L_1$ and $L_\infty$ metrics.

The reason for the worse-than-random behavior when matching the last remaining sites is that outlying zones tend to cluster together, and then all the sites in those zones are likely to have the same center as closest center. Overall, the experiments show that $\alpha=O(m)$ is a reasonable assumption.

\section{Additional results for Manhattan and Chebyshev metrics.}\label{app:moreplots}

\begin{figure}
\centering
\includegraphics[width=0.99\linewidth]{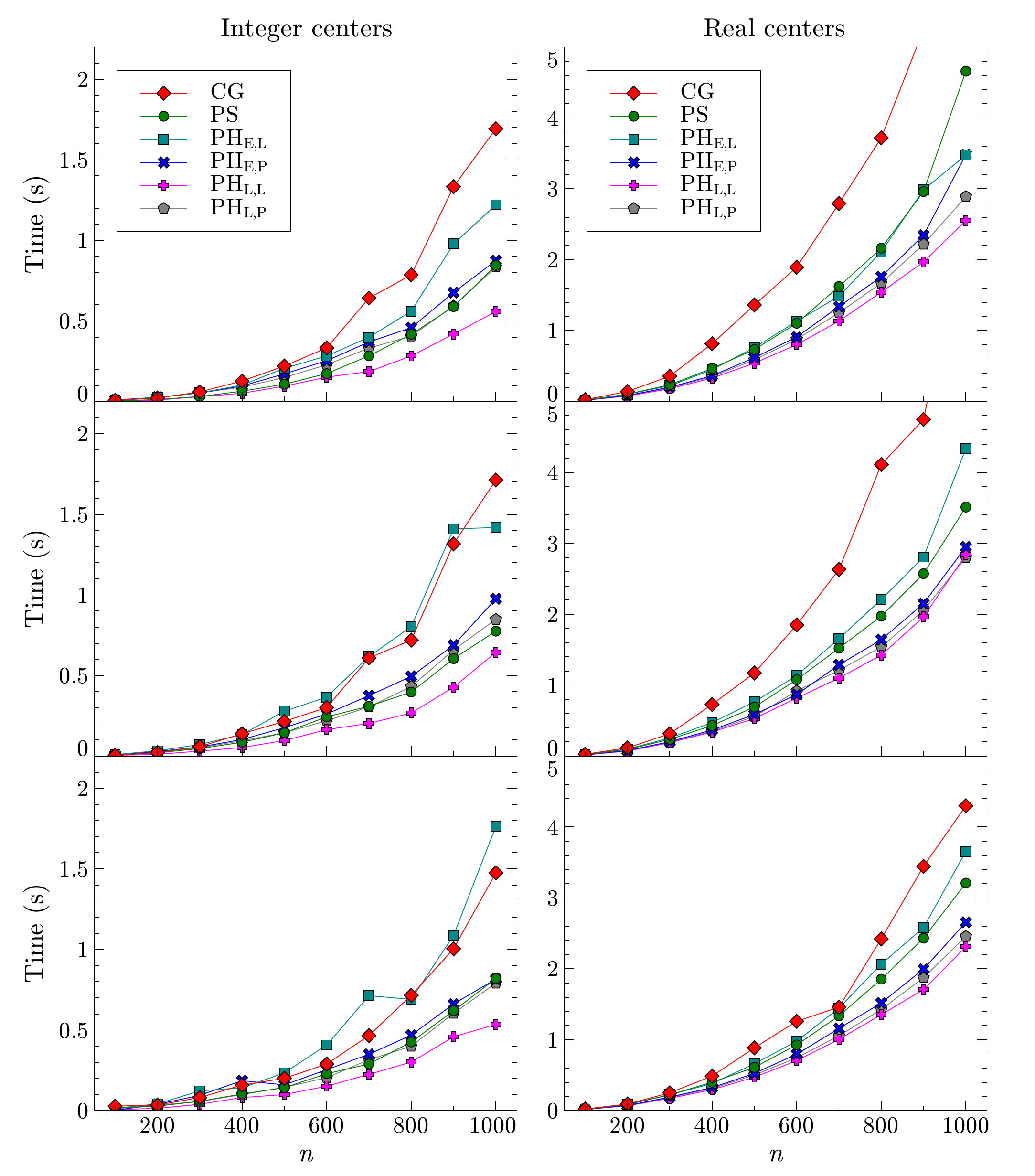}
\caption{Execution time of the various algorithms. We consider integer (left) and real (right) centers, and Euclidean (top), Manhattan (middle), and Chebyshev (bottom) metrics. For all the methods but \textit{CG}, the cutoff is $0.15$. Each data point is the average of 10 runs with $10n$ randomly distributed centers.}
\label{fig:exectimeAll}
\end{figure}

\begin{figure}
\centering
\includegraphics[width=0.99\linewidth]{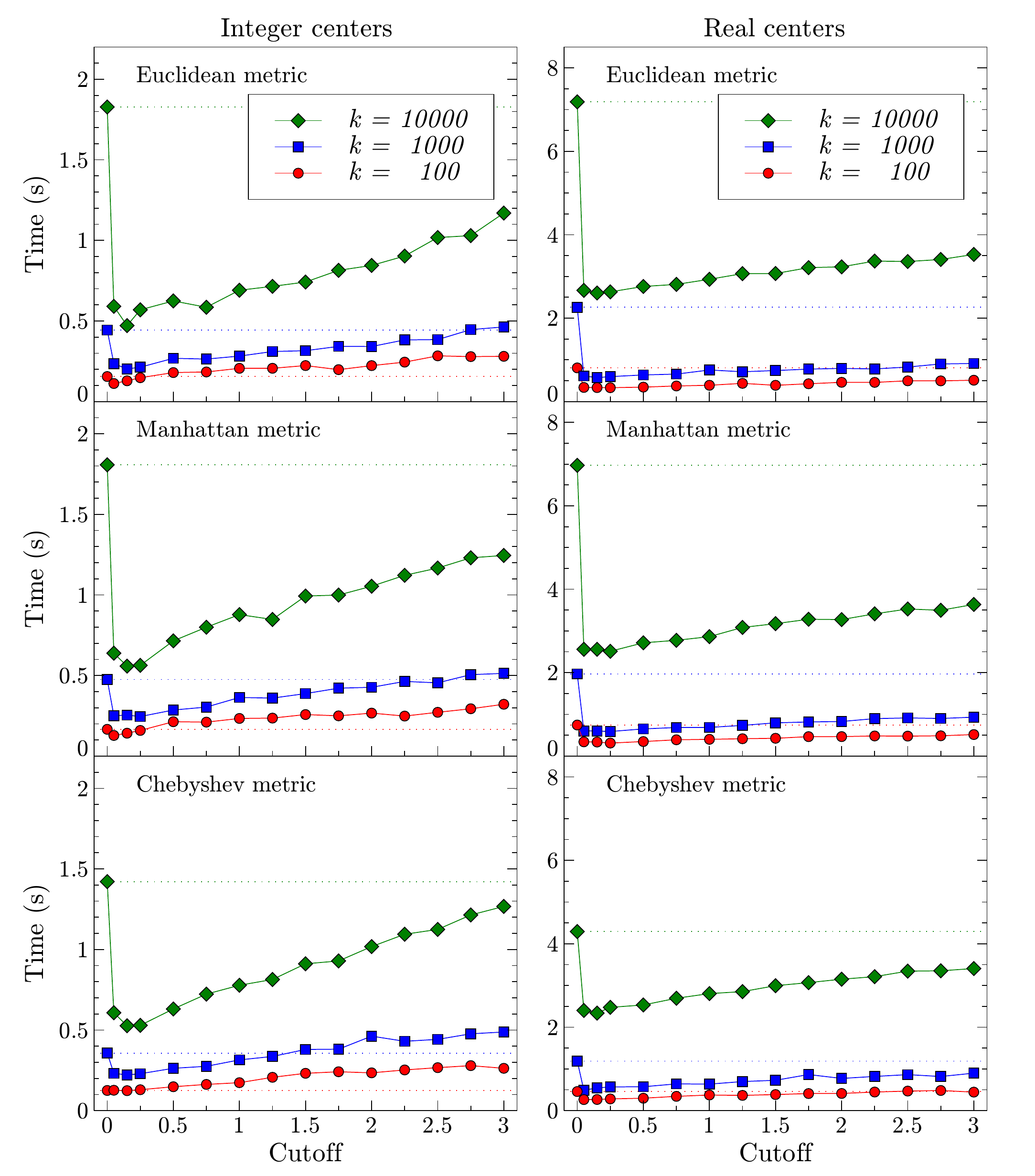}
\caption{Execution time of the circle-growing algorithm for combined with the pair heap algorithm with lazy updates and a linear search NN data structure. We consider integer (left) and real (right) centers, and Euclidean (top), Manhattan (middle), and Chebyshev (bottom) metrics. The dotted lines denote the running time of the circle-growing algorithm alone, i.e., with cutoff 0. Each data point is the average of 10 runs with randomly distributed centers, $n=1000$, and the $L_2$ metric.}
\label{fig:cutoffAll}
\end{figure}

\begin{figure}
\centering
\includegraphics[width=0.99\linewidth]{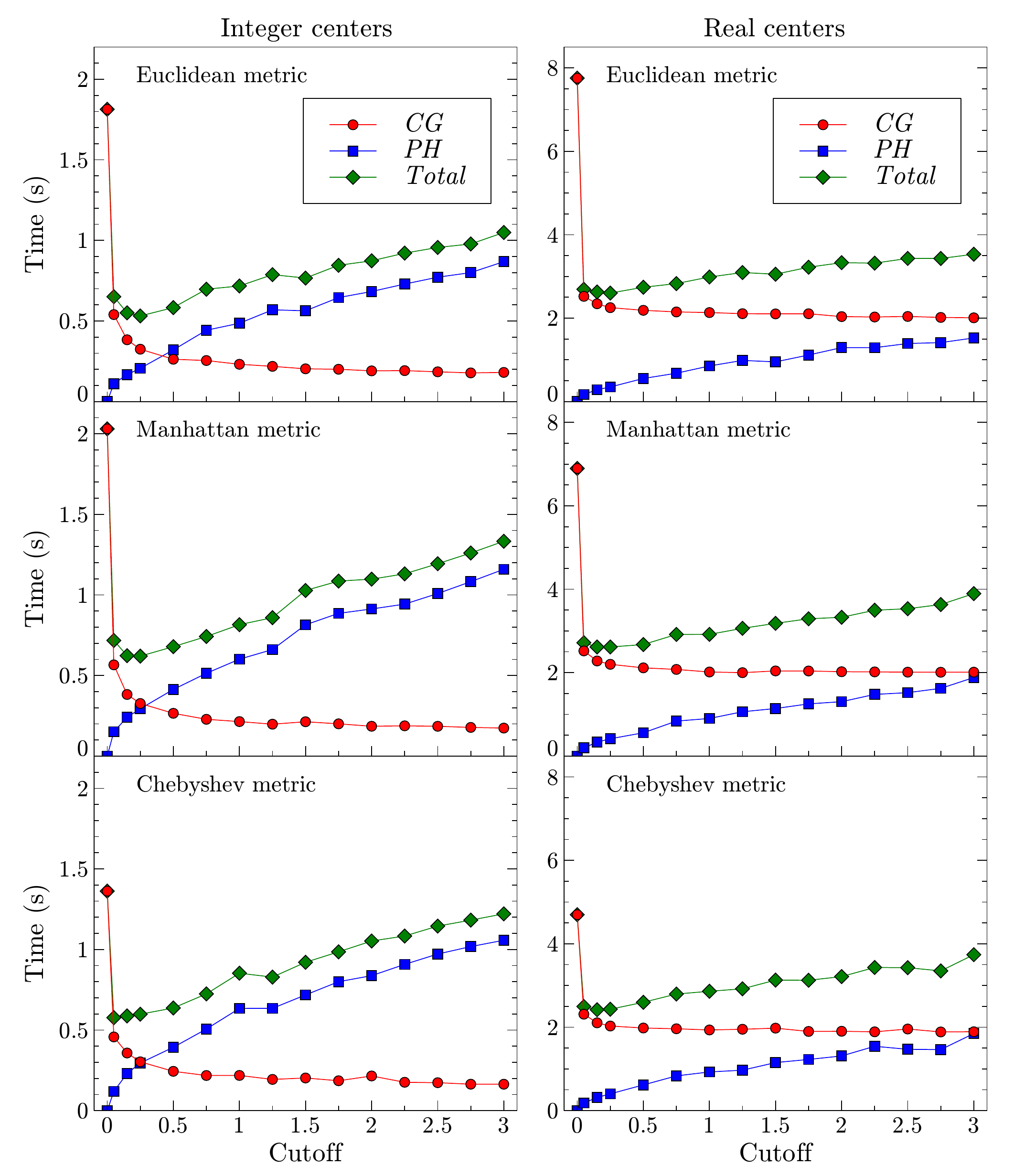}
\caption{Execution time of the circle-growing algorithm combined with the pair heap algorithm with lazy updates and a linear search NN data structure. We consider integer (left) and real (right) centers, and Euclidean (top), Manhattan (middle), and Chebyshev (bottom) metrics. In addition to the total execution time, we show the execution time spent in each algorithm. Each data point is the average of 10 runs with $n=1000$ and $10000$ randomly distributed centers.}
\label{fig:profileAll}
\end{figure}

We repeated the experiments from the experiments section for the Manhattan ($L_1$) and Chebyshev ($L_\infty$) metrics: Figure~\ref{fig:exectimeAll} extends Figure~\ref{fig:exectimeL2}, Figure~\ref{fig:cutoffAll} extends Figure~\ref{fig:cutoffL2}, and Figure~\ref{fig:profileAll} extends Figure~\ref{fig:profileL2}.

The figures in this section show that the metric used does not play a major role in the optimal cutoff nor the execution time.

\end{document}